\documentclass[10pt,a4paper,twocolumn,english,prb,aps,showpacs,floatfix,groupedaddress,superscriptaddress]{revtex4-1}

\usepackage{graphicx}
\usepackage{epsfig}
\usepackage[english]{babel}
\usepackage{amsmath}
\usepackage{amssymb}
\usepackage{amsfonts}
\usepackage{longtable}
\usepackage{dcolumn}
\usepackage{bm}
\usepackage{bbm}
\usepackage{nicefrac}
\usepackage{color,array}
\usepackage{colortbl}

\begin{document}
\title{Enhanced Perturbative Continuous Unitary Transformations}

\author{H. Krull}
\email{holger.krull@tu-dortmund.de}
\author{N. A. Drescher}
\email{nils.drescher@tu-dortmund.de}
\author{G. S. Uhrig}
\email{goetz.uhrig@tu-dortmund.de}
\affiliation{Lehrstuhl f\"{u}r Theoretische Physik I, Technische Univerit\"at Dortmund, 
Otto-Hahn Stra\ss{}e 4, 44221 Dortmund, Germany}

\date{\rm\today}

\begin{abstract}
Unitary transformations are an essential tool for the theoretical understanding
of many systems by mapping them to simpler effective models. A systematically controlled
variant to perform such a mapping is a perturbative continuous unitary transformation (pCUT)
among others. So far, this approach required an equidistant unperturbed spectrum.
Here, we pursue two goals: 
First, we extend its applicability to non-equidistant spectra with the particular
focus on an efficient derivation of the differential flow equations, which define
the enhanced perturbative continuous unitary transformation (epCUT).
Second, we show that the numerical integration of the flow equations yields a robust
scheme to extract data from the epCUT.
The method is illustrated by the perturbation of the harmonic oscillator with a quartic term and of the two-leg spin ladders in the
strong-rung-coupling limit for uniform and alternating rung couplings. The latter case
provides an example of perturbation around a non-equidistant spectrum.
\end{abstract}





\maketitle

\section{Introduction}

Quantum many-body systems with correlations are notoriously difficult to describe
theoretically. Many analytical and numerical 
tools have been developed to tackle such problems. Tools which are employed
ubiquitously are unitary transformations. Famous applications are the 
fermionic Bogoliubov transformations in the mean-field theory of superconductivity
by Bardeen, Cooper, and Schrieffer \cite{barde57} (BCS) or the bosonic Bogoliubov transformations 
arising in linear spin-wave theory of quantum antiferromagnets. \cite{auerb94}
These are exact transformations which use the algebraic properties of fermions and
bosons, respectively.
They yield diagonal Hamiltonians if they are applied to bilinear initial Hamiltonians.

Another class of unitary transformations are those which are not exact but
approximate because they rely on an expansion in a small parameter. A well-known
example is the antiferromagnetic Heisenberg exchange coupling $J$ as it is
derived from a half-filled Hubbard model with hopping $t$ and local repulsion 
$U$ implying $J=4t^2/U$ (see for instance Ref.\ \onlinecite{harri67}). Obviously, higher 
contributions $\mathcal{O}\left(t^3/U^2\right)$ are neglected, but they can also be computed systematically.
\cite{takah77,macdo88,stein97,reisc04,phill04,yang10,hamer10}

Moreover, the Hubbard model is 
not diagonalized by the transformation, but mapped to an effective spin model. This mapping
implies a simplification because the relevant part of the Hilbert space (here, spin
degrees of freedom) has been separated from the remainder (charge degrees of freedom).
The remainder does not need to be considered. It is said that
 it has been eliminated or integrated out.
Another famous example in the same line is the derivation of an 
attractive interaction between electrons from the exchange of a phonon.
This well-known step precedes the BCS theory of superconductivity.
We discuss it below in the first part of the next section because it constitutes
an excellent example that different unitary transforms yield different
effective models, even in leading order. In particular, it shows
that a continuous version generically yields
effective models with less singular coefficients as functions of the
bare parameters.
A related approach, which is not continuous but iterative, is
the projective renormalization (PRG), \cite{becke02} and it has also been
applied successfully to electron-phonon interactions. \cite{hubsc03,sykor05}

The main goals of this paper are twofold. 
First, we show how continuous unitary transformations (CUTs) can be used to perturbatively derive effective Hamilton
operators in real space. This goal has been realized for an unperturbed Hamiltonian
with equidistant spectrum  by perturbative CUT
(pCUT). \cite{uhrig98c,knett00a,knett03b} The gist of the pCUT is
 recalled in the following. The matrix elements of the effective models derived by pCUT
 have to be computed by evaluating long products of operators for various clusters.
 In this work, we enhance the applicability of such an approach
to unperturbed \emph{non}-equidistant spectra
by formulating the CUT directly in second quantization,
i.e., in the prefactors of monomials of creation and annihilation operators.
The resulting transformation
will be called enhanced perturbative CUT (epCUT) for distinction.
The approach is exemplified for a uniform and for an alternating spin ladder.
The latter has a non-equidistant spectrum if only the rung couplings
are considered.

The second main goal is to establish a robust extrapolation of the perturbative results
of the epCUT. We will show that a direct evaluation of the perturbatively established
 flow equations provides a very robust and reliable way to extrapolate the perturbative
 results. This approach will be called directly evaluated enhanced perturbative CUT (deepCUT).

The article is set up as follows. In the next section, we briefly
exemplify the versatility of continuous unitary transformations  by 
deriving the BCS electron-electron attraction. We 
introduce the perturbative CUT and the self-similar CUT (sCUT) as predecessors of the
epCUT and the deepCUT. In Sec.\ III, we introduce the harmonic oscillator with quartic perturbation and
our paradigm model, spin ladders, for which
we illustrate the general approaches. In Sec.\ IV, we derive the epCUT and develop the deepCUT
from it. Many technical aspects are discussed; a focus is the definition of simplifying rules which allow
us to compute high orders efficiently. In Sec.\ V, results of the epCUT and the deepCUT
are presented for the uniform antiferromagnetic spin ladder with $S=1/2$.
Results for the alternating spin ladder, which does not have an equidistant unperturbed spectrum,
are shown in Sec.\ VI. The article terminates by the conclusions in Sec.\ VII.

\section{Methodological Background}

The focus is here on previous variants of continuous unitary transformations
in order to show from where we start and in which respect we
go beyond presently known methods. But, there are also related approaches
such as projective renormalization, 
high-order series expansions on the basis of the linked-cluster theorem,
and the coupled-cluster method.

\subsection{Electron-electron attraction from electron-phonon interaction}

The Fr\"ohlich transformation \cite{frohl52}
eliminates phononic degrees of freedom from an electron-phonon
system in leading order of the coupling to derive an electron-electron
interaction from an electron-phonon coupling. Starting from the Hamiltonian
\begin{subequations}
\begin{align}
H &= H_{\text{D}} + H_{\text{int}},
\\
H_{\text{D}} &= \sum_{\vec{k},\sigma} \varepsilon_{\vec{k}} 
c^\dag_{\vec{k},\sigma} c^{\phantom{\dag}}_{\vec{k},\sigma} + 
\sum_{\vec{q}} \omega_{\vec{q}} b^\dag_{\vec{q}} b^{\phantom{\dag}}_{\vec{q}},
\\
H_{\text{int}} &= \sum_{\vec{k},\vec{q},\sigma} M_{\vec{q}} 
(b^{\phantom{\dag}}_{\vec{q}} + b^\dag_{-\vec{q}} )
c^\dag_{\vec{k}+\vec{q},\sigma} c^{\phantom{\dag}}_{\vec{k},\sigma},
\end{align}
\end{subequations}
this transformation generates an attractive interaction in the BCS channel 
\begin{equation}
H_\text{BCS}=\frac{1}{N}\sum_{\vec{k},\vec{k}',\sigma,\sigma'}
V_{\vec{k},\vec{k}'} \; c^\dag_{\vec{k}',\sigma'} c^\dag_{-\vec{k}',-\sigma'}
c^{\phantom{\dag}}_{-\vec{k},-\sigma} c^{\phantom{\dag}}_{\vec{k},\sigma}
\end{equation}
with the matrix element 
\begin{equation}
\label{eq:bcs_frohlich}
V_{\vec{k},\vec{k}'}^{\text{F}} = \frac{|M_{\vec{q}}|^2 \; 
\omega_{\vec{q}}}{\Delta\varepsilon^2- \omega_{\vec{q}}^2},
\end{equation}
where $\vec{q}:=\vec{k}'-\vec{k}$, $\Delta\varepsilon = \varepsilon_{\vec{k}'}- \varepsilon_{\vec{k}}$.
This explains the formation of Cooper pairs and conventional superconductivity.
It is interesting to note that in standard treatments, the interaction
is usually approximated by a constant, leaving out any discussion
of the resonance singularity in \eqref{eq:bcs_frohlich}.

It is, however, possible to achieve the elimination of the phonon degrees
of freedom by a different, continuous unitary transformation (CUT). This approach relies
on a continuously parametrized anti-Hermitian generator $\eta(\ell)=-\eta^\dag(\ell)$
of the differential unitary transformation 
\begin{equation}
\label{eq:flow}
\partial_\ell H(\ell) = [\eta(\ell),H(\ell)]
\end{equation}
of the Hamiltonian $H(\ell)$; the transformation starts at $\ell=0$
and ends\cite{wegne94,glaze93,glaze94,kehre06} at $\ell=\infty$.

One possible choice for the generator 
leading to a convergent flow \cite{wegne94} 
for $\ell\to\infty$ is $\eta^{\text{W}}:=[H_{\text{D}},H]$ where
$H_{\text{D}}$ is the diagonal part of the Hamiltonian. Integrating the
flow equation \eqref{eq:flow} from $\ell=0$ to $\infty$ yields 
for the BCS channel \cite{lenz96,kehre06} in leading order in
$M_{\vec{q}}$
\begin{equation}
\label{eq:bcs_lenz}
V_{\vec{k},\vec{k}'}^{\text{W}} = -\frac{|M_{\vec{q}}|^2 \; 
\omega_{\vec{q}}}{\Delta\varepsilon^2+ \omega_{\vec{q}}^2}.
\end{equation}
The eye-catching fact in $V_{\vec{k},\vec{k}'}^{\text{W}}$ is that it
does not have a resonant energy denominator. Hence, this result is much
smoother. In particular, it implies an attractive interaction for all parameters.

The standard BCS interaction is a constant up to
some phononic cutoff energy $\omega_{\text{Debye}}$. This result can
be derived rigorously
by a modification of the generator. In an eigen basis of $H_{\text{D}}$,
the matrix elements of $\eta^{\text{sgn}}$ are chosen 
\cite{mielk98,uhrig98c,knett00a} to be 
$\eta^{\text{sgn}}_{ij}:=\text{sgn}(E_i-E_j)H_{ij}$. Then, we find
\begin{equation}
\label{eq:bcs_mku}
V_{\vec{k},\vec{k}'}^{\text{sgn}} = -\frac{|M_{\vec{q}}|^2 }{ \omega_{\vec{q}}}
\Theta(\omega_{\vec{q}}-|\Delta\varepsilon|),
\end{equation}
where $\Theta(x)$ is the Heaviside step function. Again, there is only attractive interaction.
In addition, the interaction is only active in a restricted energy interval and zero
outside. A similar result was obtained by Mielke using a self-similar approach.
\cite{mielk97a,mielk97b}

It is very remarkable that all three approaches (\ref{eq:bcs_frohlich}), (\ref{eq:bcs_lenz}), and
(\ref{eq:bcs_mku}) are different in their outcome although 
they do the same: eliminating the linear electron-phonon coupling. We stress 
that this is not a spurious result, but relies on the fact that the unitary
transformations are indeed different even in leading order. They express
virtual processes in a different way. But, the energy-conserving
processes at $\Delta\varepsilon=0$ are the same in all three results.
This has to be so because such scattering processes can in principle be measured
which implies that they have to be independent from the chosen basis.

\subsection{Perturbative continuous unitary transformation}

We draw the readers' attention to the fact that we are dealing from now
on with CUTs with a unique reference state. This means that the ground-state is mapped by the CUT to the vacuum of excitations. In the previous section,
the mapping to effective models such as the Heisenberg exchange model or the BCS model
still left a many-body problem to be solved.

High-order series expansions have long been used to compute reliable
ground-state energies \cite{singh88b,gelfa90} and dispersions in strongly correlated systems. \cite{gelfa96,gelfa00,oitma06} No particular assumptions on the unperturbed
spectrum are required.
Ground-state energies and dispersions can be computed straightforwardly because the states are uniquely determined by their quantum number, for instance, the momentum, even before the
perturbation is switched on. This means it is sufficient to perform the perturbation 
for a one-dimensional subspace of the Hilbert space.
States of two and more particles are more subtle because their subspaces are 
extensively large. For instance, binding energies can not
be computed as series unless the binding occurs already in linear order. Generally,
unitary or orthogonal transformations must be introduced to define the perturbative approach
on large subspaces.\cite{uhrig98c,knett00a,knett03a,trebs00,oitma06}
For static ground-state properties, the coupled-cluster method \cite{bisho98b} represents
also a powerful means to learn systematically from clusters of finite size 
about the physics of the thermodynamic limit.

The perturbative CUT (pCUT) was the first approach to systematically address
many-particle states. \cite{uhrig98c,knett00a} 
Its starting point is a Hamiltonian which can be written in the form
\begin{align}
 H(0)&=H_0+x\sum_{m=-N}^N T_m,
\end{align}
where $H_0$ is the unperturbed Hamiltonian with an equidistant spectrum
as additional assumption.
For simplicity, we set its energy spacing to unity. Each energy quantum can be
seen as an elementary excitation, a quasiparticle, so that $H_0$ counts the number
of quasiparticles up to an irrelevant constant offset.
The expansion parameter is $x$ and the terms in the perturbation are split according to
their effect on the quasiparticle number $H_0$: The terms in $T_m$ increase the
number of energy quanta by $m$. Obviously, $T_{-m}=T_m^\dag$ holds. Generically,
there is an upper bound $N\ge |m|$ to the change of energy quanta. 
In pCUT, the parametrization
\begin{align}
H(\ell)&=H_0+\sum\limits_{k=1}^\infty x^k \sum_{\text{dim}(\vec{m})=k} F(\ell;\vec m) T_{\vec m}
\end{align}
is used as an ansatz for the flowing Hamiltonian with coefficients $F(\ell;\vec m)$. The components of the vector $\vec{m}$ take the values $-N,-N+1,\ldots,N-1,N$;
the vector has the dimension $\text{dim}(\vec{m})=k$. The notation $T(\vec{m})$ stands for the product
$T_{m_1}T_{m_2}\ldots T_{m_{k-1}}T_{m_k}$. \cite{knett00a}
Choosing
\begin{align}
 \eta^{\text{pc}}(\ell)=\sum_{k=1}^\infty x^k\sum_{\text{dim}(\vec{m})=k} \text{sgn}(M(\vec m)) F(\ell;\vec m) T(\vec m)
\end{align}
the flow equation \eqref{eq:flow} generates a hierarchy of differential equations in powers of $x$ for the coefficients $F(\ell;\vec m)$.
In each finite order, the differential equations are closed and can be solved 
by computer aided analytics. Eventually, one obtains the general expansion
for an effective Hamiltonian which conserves the number of elementary excitations
\begin{equation}
\label{eq:pCUT_fund}
H_{\text{eff}} = H_0+\sum_{k=1}^\infty x^k
\sum_{\text{dim}(\vec{m})=k,M(\vec{m})=0} C(\vec{m}) T(\vec{m}).
\end{equation}
 The renormalized coefficients $C(\vec{m})=F(\infty;\vec m)$ are
fractions (without imaginary part) and the conservation
of the number of quasiparticles is implied by the cross sum 
$M(\vec{m})=0$ where $M(\vec{m})=\sum_{j=1}^k m_j$.

The result \eqref{eq:pCUT_fund} is very general; to put it to practical use its
irreducible effect on zero, one, two, and more quasiparticles is computed.
In this way, the effective Hamiltonian is obtained in second quantized form \cite{knett03b}.
Remarkable achievements of this approach are a quantitative understanding
of inelastic scattering in spin ladders 
\cite{windt01,schmi01,knett01b,schmi05b,notbo07,hong10b},
of spectral densities in spin chains \cite{schmi04a}, of
excitations in the Kitaev model \cite{schmi08a,vidal08}, 
of excitations in the toric code \cite{vidal09a,vidal09b,dusue11},
and of the ionic Hubbard model \cite{hafez10}
to name a few extended systems where the ground-state is described
as a vacuum of excitations.

Conceptually, 
the most significant achievement of pCUT is that whole subspaces are treated perturbatively. 
The generality of the pCUT result \eqref{eq:pCUT_fund} is surely one of its
advantages. The fact that it can only deal with equidistant spectra is a certain caveat,
not shared by the high-order series expansions described in Ref.\ \onlinecite{oitma06}.
Another caveat is that the approach does not allow for modifications of the generator.

\subsection{Self-Similar Continuous Unitary Transformations}

One way to circumvent the above-mentioned restrictions concerning the unperturbed
spectrum and the choice of the generator is to pass from a perturbative evaluation
to a self-similar one. The approach follows a straightforward strategy. One chooses
a set of operators, which serves as a basis. The Hamiltonian and the generator 
are described as linear combinations of these operators. By commuting Hamiltonian
and generator and re-expanding the result in the same operator basis, the flow
equation \eqref{eq:flow} induces a differential equation system (DES) in the coefficients
of the basis operators. A more detailed description follows in Sec.\ III below. 
We stress that in the latter step a certain truncation is
required. \cite{dresc11} Unless the set of operators is closed under commutation, 
the commutator $[\eta,H]$ comprises terms which can not be expanded exactly in the 
operator basis. Thus, this step generically requires an approximation. The Hamiltonian 
is kept in a self-similar form defined by the selected operator basis.

Depending on the system, the truncating approximation can be controlled by a small parameter
\cite{mielk97a,mielk97b,kehre06}
or by the spatial locality of the selected set of operators.\cite{reisc04,Fischer2010}
We stress that in the sCUT approach, the choice of the operator basis and of the generator
uniquely defines the DES of the flow equation.
Clearly, the advantage of the sCUT over the pCUT is its larger versatility.
Yet, it is less general in the sense that the flow equation has to be solved
for each model and each operator basis anew. 

To derive a systematic perturbative expansion by sCUT is not an obvious step.
It is one of our two main goals to show how this can be done and
how it can be done efficiently. Thus, the derivations and considerations
in Sec.\ IV are based on the sCUT approach and combine it with 
a perturbative expansion in order to reach the enhanced perturbative CUT.

\section{Models}

We illustrate our approach by applying it to two models which are introduced
below. The first is a zero-dimensional, perturbed harmonic oscillator and
it is chosen for its simplicity. The second are one-dimensional spin ladders
which represent well-understood extended models.

\subsection{Harmonic oscillator with quartic perturbation}
\label{struct:model_toy}
We analyze the perturbed harmonic oscillator
\begin{subequations}
\label{eq:H_toy}
\begin{align}
 H_{}&= \epsilon_0 + \omega {b^{\dagger}}{}{b^{}} +x\cdot H_1
\end{align}
 with ground-state energy $\epsilon_0$, frequency $\omega_0>0$, and bosonic creation and annihilation operators ${b^{\dagger}}{},{b^{}}$. It is perturbed by 
\begin{align}
 H_1 &= {b^{\dagger}}{}^4+{b^{}}^4+\tilde\epsilon + \tilde\omega{b^{\dagger}}{}{b^{}} + U{b^{\dagger}}{}{b^{\dagger}}{}{b^{}}{b^{}},
 \end{align}
\end{subequations}
controlled by the expansion  parameter $x$.
The perturbation includes a ground-state shift $\tilde\epsilon$, a frequency shift $\tilde\omega$, and a density-density repulsion $U$. In order that $H$ is bounded from below for
$x\in[0,\infty)$, we require $H_1$ to be positive. Using the Ger\v{s}gorin circle theorem 
\cite{Gervsgorin1931}
to the diagonal elements $e_n$ of $H_1$ in the basis of oscillator eigenstates $\{|n\rangle\}$, 
all eigenvalues are positive if 
\begin{subequations}
\begin{align}
e_n&=\left\langle n\right|H_1\left|n\right\rangle=\epsilon + n\tilde\omega + n(n-1)U\\
   &\overset{!}{>} \left|\left\langle n+4\right|H_1\left|n\right\rangle\right| + \left|\left\langle n-4\right|H_1\left|n\right\rangle\right|
\end{align}
\end{subequations}
holds. The second matrix element occurs only for $n\ge 4$ and can be estimated by $\left\langle n+4\right|H_1\left|n\right\rangle>\left\langle n-4\right|H_1\left|n\right\rangle> 0$.
The resulting final inequality
\begin{align}
e_n^2{>} 4(n+4)(n+3)(n+2)(n+1)
\end{align}
is satisfied for
 $\tilde\epsilon=10$, $\tilde\omega=12$, and $U=2$.

\subsection{Spin-$\frac{1}{2}$ Heisenberg ladder}

\begin{figure}
 \includegraphics[width=\columnwidth]{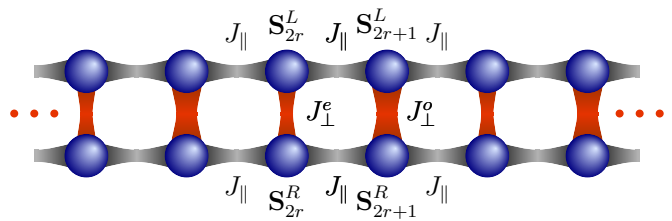}
 \caption{(Color online) Schematic representation of the uniform (alternating)
 	$S=\nicefrac{1}{2}$ Heisenberg ladder in the thermodynamic limit.}
 \label{img:ladder}
\end{figure}

To illustrate the performance
of the (de)epCUT, we consider the S=$\frac{1}{2}$ antiferromagnetic two-leg
Heisenberg ladder (uniform spin ladder) and an extension with an alternating rung coupling 
(alternating spin ladder) as testing ground (see Fig.\ \ref{img:ladder}).  
The Hamiltonian reads as
\begin{subequations}
\begin{align} H&=J_{\bot}^{\text{e}}H_{\bot}^{\text{e}}+J_{\bot}^{\text{o}}H_{\bot}^{\text{o}}+J_{\parallel}H_{\parallel} \\
H_{\bot}^{\text{e}}&=\sum\limits_{r=0}^{L/2-1}
\textbf{S}_{2r}^{\text{L}}\cdot\textbf{S}_{2r}^{\text{R}}\\
H_{\bot}^{\text{o}}&=\sum\limits_{r=0}^{L/2-1}
\textbf{S}_{2r+1}^{\text{L}}\cdot\textbf{S}_{2r+1}^{\text{R}}\\
H_{\parallel}&=\sum\limits_{r=0}^{L-1}\left(\textbf{S}_{r}^{\text{L}}\cdot \textbf{S}_{r+1}^{\text{L}}+\textbf{S}_{r}^{\text{R}}\cdot \textbf{S}_{r+1}^{\text{R}}\right)\,\text{,}
  \label{eq:spinhammi}
\end{align}
\end{subequations}
where $r\in \mathbb{Z}$. The rung number is denoted by $r$ and the legs by L and R. 
We define the ratio between the leg coupling $J_{\parallel}$ and the even rung coupling $J_{\bot}^{\text{e}}$ as relative leg coupling $x:=\nicefrac{J_{\parallel}}{J_{\bot}^{\text{e}}}$ and the ratio between the odd rung coupling $J_{\bot}^{\text{o}}$ and the even rung coupling $J_{\bot}^{\text{e}}$ as $y:=\nicefrac{J_{\bot}^{\text{o}}}{J_{\bot}^{\text{e}}}$.

In the limit of $J_{\bot}^{\text{e}}=J_{\bot}^{\text{o}}$, i.e., $y=1$, the Hamiltonian describes the uniform
spin ladder. This model has been subject of intensive studies (see Refs.\ \onlinecite{johns00a,Dagotto2002,schmi05b}
and references therein). Thus, it constitutes a suitable reference model to test the epCUT. 
It has
been investigated by several different methods, such as density matrix renormalization, \cite{nunne02,schmi12} 
exact diagonalization, \cite{brehm99} continuum field theory,
\cite{Shelton96,schul96a} quantum Monte Carlo, \cite{greve96} high order series expansions, \cite{Zheng1998,oitma96b}
including methods based on CUTs, such as sCUT (Refs.\ \onlinecite{Fischer2010,Duffe10,Duffe11}) and pCUT. \cite{knett01b,knett03b,schmi04b,schmi05b} 
If the results of the epCUT agree with these data, the efficiency of the epCUT for the expansion around an unperturbed equidistant spectrum is verified. 

To illustrate that the epCUT represents an advancement compared to pCUT we will show
results for the alternating spin ladder as well. This system does not have an equidistant spectrum
because the rung couplings are not equal $J_{\bot}^{\text{e}}\neq J_{\bot}^{\text{o}}$.
Hence it cannot be dealt with by pCUT. Without loss of generality we consider
$J_{\bot}^{\text{o}} > J_{\bot}^{\text{e}}$ implying $y> 1$.

For the alternating spin ladder we expect a lowering of the ground-state energy upon rising $y$
because the expectation value of $\langle\textbf{S}_{r}^{\text{L}}\cdot\textbf{S}_{r}^{\text{R}}\rangle$
is negative. The unit cell includes two rungs which implies two triplon branches in the Brillouin zone (BZ). For $y=1$ the branches meet at the BZ boundary ($k=\pm\nicefrac{\pi}{2}$). For $y>1$ a band gap of the order
 of $|y-1|$ opens at $k=\pm\nicefrac{\pi}{2}$ separating the two band.

To define a starting point for the CUT the bond operator representation \cite{Chubukov89,sachd90} is used. A possible eigen basis of the local operators $\textbf{S}_{r}^{\text{L}},\textbf{S}_{r}^{\text{R}}$ is given by the singlet state
\begin{align}
\left|s\right\rangle=\frac{1}{\sqrt{2}}(\left|\uparrow \downarrow\right\rangle-\left|\downarrow \uparrow\right\rangle)
\end{align}
and the three triplet states
\begin{subequations}
\begin{align}
t^\dagger_{x}\left|s\right\rangle&:=\left|t_{x}\right\rangle=\frac{-1}{\sqrt{2}}\left(\left|\uparrow \uparrow\right\rangle-\left|\downarrow \downarrow\right\rangle\right)\\
t^\dagger_{y}\left|s\right\rangle&:=\left|t_{y}\right\rangle=\frac{i}{\sqrt{2}}\left(\left|\uparrow \uparrow\right\rangle+\left|\downarrow \downarrow\right\rangle\right)\\
t^\dagger_{z}\left|s\right\rangle&:=\left|t_{z}\right\rangle=\frac{1}{\sqrt{2}}\left(\left|\uparrow \downarrow\right\rangle+\left|\downarrow \uparrow\right\rangle\right)\text{.}
\end{align}
\end{subequations}
For $x=0$, the ground-state of the system is given by 
\begin{align}
\left|0\right\rangle:=\prod_{r}\left|s\right\rangle_r\,\text{.}
\end{align}
This vacuum of triplets serves as our reference state. The local operators $t^\dagger_{x,r},t^\dagger_{y,r}$ and $t^\dagger_{z,r}$ ($t^{\phantom{\dagger}}_{x,r},t^{\phantom{\dagger}}_{y,r}$ and $t^{\phantom{\dagger}}_{z,r}$) create (annihilate) an excitation on rung $r$. 
They satisfy the hardcore-boson commutation relation
\begin{align}
[t^{\phantom{\dagger}}_{\alpha,r},t^\dagger_{\beta,s}]=\delta_{r,s}\big(\delta_{\alpha,\beta}-t^\dagger_{\beta,r}t^{\phantom{\dagger}}_{\alpha,r}-\delta_{\alpha,\beta}\sum_{\gamma}t^\dagger_{\gamma,r}t^{\phantom{\dagger}}_{\gamma,r}\big)\text{.}
\end{align}
The elementary magnetic excitations (S=1), known as triplons \cite{schmi03c,schmi05b}
can be continuously linked to the local triplets.

Represented in second quantization in terms of the triplon creation and annihilation operators the Hamiltonian reads
\begin{align}
\frac{H}{J_{\bot}^{\text{e}}}&=H_{\bot}^{\text{e}}+yH_{\bot}^{\text{o}}+xH_{\parallel}\,\text{,}\label{eq:triplonhammi}
\end{align}
where
\begin{subequations}
\begin{align}
H_{\bot}^{\text{e}}&=-\frac{3}{4}\sum_{\phantom{1}r=2a\phantom{+}}\mathbbm{1}+\sum_{\phantom{1}r=2a\phantom{+}}t^\dagger_{\alpha,r}t^{\phantom{\dagger}}_{\alpha,r}\\
H_{\bot}^{\text{o}}&=-\frac{3}{4}\sum_{r=2a+1}\mathbbm{1}+\sum_{r=2a+1}t^\dagger_{\alpha,r}t^{\phantom{\dagger}}_{\alpha,r} \\
H_{\parallel}&=\phantom{+}\frac{1}{2}\sum_{\phantom{\neq,}r,\alpha\phantom{,\beta}}\left(t^\dagger_{\alpha,r}t^{\phantom{\dagger}}_{\alpha,r+1}+t^\dagger_{\alpha,r+1}t^{\phantom{\dagger}}_{\alpha,r}\right) \\
&\quad+\frac{1}{2}\sum_{\phantom{,}r,\alpha\neq\beta\phantom{,}}t^\dagger_{\alpha,r}t^\dagger_{\beta,r+1}t^{\phantom{\dagger}}_{\beta,r}t^{\phantom{\dagger}}_{\alpha,r+1} \\
&\quad-\frac{1}{2}\sum_{\phantom{,}r,\alpha\neq\beta\phantom{,}}t^\dagger_{\alpha,r}t^\dagger_{\alpha,r+1}t^{\phantom{\dagger}}_{\beta,r}t^{\phantom{\dagger}}_{\beta,r+1} \\
&\quad+\frac{1}{2}\sum_{\phantom{\neq,}r,\alpha\phantom{,\beta}}\left(t^\dagger_{\alpha,r}t^\dagger_{\alpha,r+1}+t^{\phantom{\dagger}}_{\alpha,r}t^{\phantom{\dagger}}_{\alpha,r+1}\right),
\end{align}
\end{subequations}
where $a, r\in \mathbb{Z}$.
This form of the Hamiltonian enters all the calculation described in the following.

\section{Derivation}

\subsection{Flow equation in second quantization}
\label{struct:flow_Hamiltonian}

Similar to the implementation of previous CUT methods \cite{wegne94,reisc04,knett03b,Fischer2010}, 
we formulate the flow equation \eqref{eq:flow} for the coefficients of the monomials $\{A_i\}$ of operators in second quantization. The Hamiltonian is parametrized by
\begin{align}
 H(\ell)=\sum\limits_i h_i(\ell)A_i
\end{align}
with the $\ell$-dependent coefficients $h_i(\ell)$. The generator reads
\begin{align}
 \eta(\ell)=\sum\limits_i \eta_i(\ell)A_i := \sum\limits_i h_i(\ell)\hat \eta [A_i]
\end{align}
with $\hat \eta$ being a superoperator denoting the application of a particular generator scheme such as those discussed in Ref.\ \onlinecite{Fischer2010}.
Expanded in the operator basis $\{A_i\}$ the flow equation \eqref{eq:flow} reads
\begin{align}
 \sum\limits_i \partial_\ell h_i(\ell) A_i = \sum\limits_{jk} h_j(\ell) h_k(\ell) \left[ \hat\eta [A_j],A_k \right] .
\end{align}
Comparing the coefficients of different monomials, the flow equation \eqref{eq:flow}
becomes equivalent to a set of ordinary differential equations for the coefficients $h_i(\ell)$
\begin{subequations}
\label{eq:flow_combined}
\begin{align}
\partial_\ell h_i(\ell) = \sum\limits_{jk} D_{ijk} h_j(\ell)h_k(\ell) .
\label{eq:flow_numeric}
\end{align}
The commutator relations between the basis operators are encoded in the coefficients $D_{ijk}$ of the bilinear differential equation system (DES). These coefficients $D_{ijk}$ are in general complex numbers.
For the spin ladders under study they are given by integers or fractional numbers. We call a single $D_{ijk}$ a `contribution' of the DES. The contributions are obtained from
\begin{align}
 \left[ \hat\eta [A_j],A_k \right] = \sum\limits_i D_{ijk} A_i
\label{eq:flow_algeraic}
\end{align}
\end{subequations}
by comparing the coefficients of the expansion of the commutator monomial by monomial.

In this way, the problem of solving the flow equation is transformed into the algebraic problem of
calculating the coefficients of the DES \eqref{eq:flow_algeraic} and 
of the subsequent numerical solution of Eq.\ \eqref{eq:flow_numeric}.

\subsection{Perturbative expansion of the flow equation}

Here we consider  the perturbative solution of the flow equation
which yields the resulting effective Hamiltonian in the form
of a perturbative series. Hence this solution generalizes the
established pCUT approach \cite{uhrig98c,knett00a}.
To this end, we decompose the initial Hamiltonian
\begin{align}
 H=H_0+x V \label{eq:H-initial}.
\end{align} 
into an unperturbed part $H_0$ and a perturbation $V$.
In contrast to pCUT \cite{knett00a}, we do not require the unperturbed part to have an equidistant spectrum. 
The formalism is very general and does not require further restrictions. In order to be able to
guarantee that a finite order in the expansion parameter requires only to deal with a finite
number of terms we assume either that the local Hilbert space at a given site is finite dimensional
\emph{or} that $H_0$ is a sum of local terms which are bilinear in bosonic or fermionic variables. We will see that the method works best for a (block-)diagonal $H_0$. 
These conditions are sufficient, but not
necessary for epCUT to work. It is beyond the scope of the present work to fully elucidate
the marginal cases where epCUT is impossible.

We aim at the perturbation series up to and including  order $n$ in $x$. 
Thus we expand the flowing Hamiltonian
\begin{align}
 H(\ell)= \sum\limits_{m=0}^n H^{(m)},\quad H^{(m)}\propto x^m \label{eq:decompH}
\end{align}
into terms of order $x^m$ up to $m\le n$.
Expanding the $H^{(m)}$ in the operator basis $\{A_i\}$ we perform
the expansion in powers of $x$ by expanding the coefficient $h_i(\ell)$ of $A_i$
\begin{align}
\label{eq:taylor-h}
 h_i(\ell)=\sum\limits_{m=0}^n x^m f_i^{(m)}(\ell). 
\end{align}
At $l=0$, the initial values $f_i^{(m)}(0)$ are fixed by the initial Hamiltonian \eqref{eq:H-initial}
and its representation in terms of the $\{A_i\}$. 
Applying \eqref{eq:taylor-h} to Eq.\ \eqref{eq:flow_numeric} one obtains
\begin{align}
 \partial_\ell \sum\limits_{m=0}^n x^m f_i^{(m)}(\ell) = \sum\limits_{j,k} D_{ijk} \sum\limits_{p,q=0}^{n}x^{p+q}f_j^{(p)}(\ell)f_k^{(q)}(\ell).
\end{align}
For the prefactors of $x^m$ this implies
\begin{align}
 \partial_\ell f_i^{(m)}(\ell) = \sum\limits_{j,k}\sum\limits_{p+q=m} D_{ijk}f_j^{(p)}(\ell) f_k^{(q)}(\ell).
\label{eq:flow_perturbative}
\end{align}

We stress that the contributions $D_{ijk}$ do not depend on the order $m$ of the coefficients, but only on the algebraic relations between the corresponding monomials. Hence they need to be calculated only
once. Moreover, Eq. \eqref{eq:flow_perturbative} defines a hierarchy between the coefficients
because $f_i^{(m)}(\ell)$ is influenced only by coefficients of the same order $m$ or lower, but not by coefficients of higher orders.

\subsection{Motivating Example}
\label{struct:example_toy}

\begin{table*}
 \begin{tabular}{|ccrclccc|}
\hline
$i$ 	&	 $A_i$						&	$h_i(0)$	&\hspace{0.8cm}	&	$h_i(\infty)$	&	${O_\text{min}^{}}$	&	${O_\text{max}^{0 \text{QP}}}$	&	${O_\text{max}^{1 \text{QP}}}$	\\
\hline
0	&		$	\mathbbm{1}			$	&	$\epsilon_0$	&	&	$\epsilon_0+\tilde\epsilon x-\nicefrac{6}{\omega_0}x^2$	&	0	&		2		&		2		\\
1	&		$	{b^{\dagger}}{}{b^{}}			$	&	$\omega_0+\tilde\omega x$	&	&	$\omega_0+\tilde\omega x-\nicefrac{24}{\omega_0}x^2$	&	0	&		0		&		2		\\
2	&		$	{b^{\dagger}}{}^4+{b^{}}^4$	&	$x$	&	&	0	&	1	&		1		&		1		\\			
3	&		$	{b^{\dagger}}{}{b^{\dagger}}{}{b^{}}{b^{}}			$	&	$Ux$	&	&	$Ux-\nicefrac{18}{\omega_0}x^2$	&	1	&	\cellcolor[gray]{0.8} 	-1		&	\cellcolor[gray]{0.6} -1		\\	
4	&		$	{b^{\dagger}}{}^3{b^{}}^3$				&	0	&	&	$-\nicefrac{4}{\omega_0}x^2$	&	2	&	\cellcolor[gray]{0.8} 	-		&	\cellcolor[gray]{0.6} 	-	\\	
5	&		$	{b^{\dagger}}{}^5{b^{}}^3 + {b^{\dagger}}{}^3{b^{}}^5$	&	0	&	&	0	&	2	&	\cellcolor[gray]{0.8} 	-		&	 \cellcolor[gray]{0.6} 	-		\\			
\hline
\end{tabular}
\caption{Basis operators $A_i$ (simple combinations of monomials obeying hermiticity) occuring in
 a second order epCUT for the perturbed harmonic oscillator using the particle conserving generator scheme ${\widehat \eta_\text{pc}}$. The third column shows the initial coefficients $h_i(\ell=0)$, the fourth the final renormalized coefficients $h_i(\ell=\infty)$. The minimum order ${O_\text{min}^{}}$ is the leading order of the considered operator; 
 ${O_\text{max}^{0 \text{QP}}}$ is the highest relevant order of the coefficient
 for computing the ground-state energy, and ${O_\text{max}^{1 \text{QP}}}$ is the highest relevant order
 for computing the excitation energy (c.f. Sec. \ref{struct:omax}).
 The terms marked in light gray are irrelevant for the computation of the
 ground-state energy in second order; the terms in dark gray are irrelevant if the excitation energy is computed. If a term can not influence the targeted quantities at all, it has no maximal order (symbolized by a dash).
\label{tab:toy_list_2nd_order}}
\end{table*}

\begin{table}
\begin{tabular}{rrrrrrrrrrr}
\hline
$i$	&		2		&		0		&		1		&	\cellcolor[gray]{0.6} 	3		&	\cellcolor[gray]{0.6} 	4		&		2		&	\cellcolor[gray]{0.6} 	5		&	\cellcolor[gray]{0.6} 	5		&	\\	
$j$	&		2		&		2		&		2		&		2		&		2		&		2		&		2		&	\cellcolor[gray]{0.6} 	5		&	\\	
$k$	&		1		&		2		&		2		&		2		&		2		&	\cellcolor[gray]{0.6} 	3		&	\cellcolor[gray]{0.6} 	3		&		1		&	\\	\hline
$D_{ijk}$	&		-4		&		-48		&	\cellcolor[gray]{0.8} 	-192		&	\cellcolor[gray]{0.6} 	-144		&	\cellcolor[gray]{0.6} 	-32		&	\cellcolor[gray]{0.6} 	-12		&	\cellcolor[gray]{0.6} 	-8		&						
\cellcolor[gray]{0.6} 	-4		&	\\	\hline
\end{tabular}
\caption{Non-vanishing contributions $D_{ijk}$ to the differential equation system (DES) of the perturbed harmonic oscillator in the particle conserving generator scheme. Operators and contributions marked in light gray are irrelevant for the computation of the ground-state energy in second order; those in dark gray are irrelevant for the first excitation. 
\label{tab:toy_diffeq_2nd_order}}
\end{table}

As simple illustration, we analyze the perturbed harmonic oscillator in Eq.\ \eqref{eq:H_toy}
using the particle-conserving generator scheme $\hat\eta_{\text{pc}}$. For order zero, we parametrize the prefactors of the unperturbed parts by the flow parameter $\ell$ leading to
\begin{align}
 H^{(0)}(\ell)
  &=f_0^{(0)}(\ell)\underbrace{\mathbbm{1}}_{A_0}+f_1^{(0)}(\ell)\underbrace{{b^{\dagger}}{}{b^{}}}_{A_1}
\end{align}
with the initial conditions $f_0^{(0)}(0)=\epsilon_0$ and $f_1^{(0)}(0)=\omega_0$. 
(The operators $A_i$ are also listed in Tab.\ \ref{tab:toy_list_2nd_order}.) 
None of these terms  contributes to the generator. Hence the coefficients
stay constant in order zero.

In linear order, two additional terms $A_2$ and $A_3$ occur
\begin{align}
\begin{split}
H^{(1)}(\ell)&= xf_0^{(1)}(\ell)\underbrace{\mathbbm{1}}_{A_0}+ xf_1^{(1)}(\ell)\underbrace{{b^{\dagger}}{}{b^{}}}_{A_1}\\
 &+ xf_2^{(1)}(\ell)\underbrace{\left({b^{\dagger}}{}^4+{b^{}}^4\right)}_{A_2}+ 
 x f_3^{(1)}(\ell)\underbrace{{b^{\dagger}}{}{b^{\dagger}}{}{b^{}}{b^{}}}_{A_3}
\end{split}
\end{align}
with the initial conditions  $f_0^{(1)}(0)=\tilde \epsilon$, $f_1^{(1)}(0)=\tilde \omega$, $f^{(1)}_2(0)=1$ and $f^{(1)}_3(0)=U$. The third term contributes to the generator
\begin{align}
 \eta^{(1)}(\ell)=xf_2^{(1)}(\ell)\left({b^{\dagger}}{}^4-{b^{}}^4\right)\ .
\end{align}
Because $\eta = O(x)$ the derivative in linear order reads as
\begin{subequations}
\begin{align}
 \partial_\ell H^{(1)}(\ell) &=[\eta^{(1)}(\ell),H^{(0)}(\ell)]\\
 &= xf^{(1)}_2(\ell)f^{(0)}_1(\ell)\left[\hat\eta_{\text{pc}} A_2, A_1\right]\\
 &= -4xf^{(1)}_2(\ell)f^{(0)}_1(\ell)\underbrace{\left({b^{\dagger}}{}^4+{b^{}}^4\right)}_{A_2}.
\end{align}
\end{subequations}
By comparing coefficients, one identifies the contribution $D_{221}$ 
to  $f^{(1)}_2$:
\begin{align}
 \partial_\ell f^{(1)}_2(\ell)=-4\omega_0 f^{(1)}_2(\ell)
\end{align}
with the initial condition $f^{(1)}_2(0)=1$ and the solution
\begin{align}
  f^{(1)}_2(\ell)=e^{-4\omega_0\ell}.
\end{align}
All other first-order coefficients retain their initial values.

The initial Hamiltonian does not comprise second-order terms.
Such terms  arise due to commutation of terms of lower order. The two relevant combinations are
\begin{align}
 \partial_\ell H^{(2)}(\ell) &= [\eta^{(1)}(\ell),H^{(1)}(\ell)]+[\eta^{(2)}(\ell),H^{(0)}(\ell)].
\end{align}
The first one reads as
\begin{align}
\nonumber
 [\eta^{(1)}(\ell),H^{(1)}(\ell)] =&x^2f^{(1)}_2(\ell)f^{(1)}_1(\ell)\left[\hat\eta_{\text{pc}} A_2, A_1\right]\\ 
 \nonumber
 +&x^2f^{(1)}_2(\ell)f^{(1)}_2(\ell)\left[\hat\eta_{\text{pc}} A_2, A_2\right]\\
 +&x^2f^{(1)}_2(\ell)f^{(1)}_3(\ell)\left[\hat\eta_{\text{pc}} A_2, A_3\right].
\end{align}
To represent the right-hand side, two additional terms $A_4$ and $A_5$ are required
(see Table\ \ref{tab:toy_list_2nd_order}):
\begin{subequations}
\begin{align}
 \left[\hat\eta_{\text{pc}} A_2, A_1\right]&=-\phantom{0}4A_2,\\
 \left[\hat\eta_{\text{pc}} A_2, A_2\right]&=-48A_0-192A_1-144A_3-32A_4\label{eq:toy_de_identity_2},\\
 \left[\hat\eta_{\text{pc}} A_2, A_3\right]&=-12A_2-\phantom{00}8A_5  
\end{align}
\end{subequations}
with vanishing initial values $f^{(2)}_{4,5}(\ell)=0$.

In Table\ \ref{tab:toy_diffeq_2nd_order}, we summarize the explicit results for the contributions to the differential equation system.
Here, we focus on the second-order correction to the identity operator $A_0$, i.e., on the ground-state energy  $E_0^{(2)}=f_0^{(0)}(\infty)+xf_0^{(1)}(\infty)+x^2f_0^{(2)}(\infty)$. Because the only second order contribution to $A_0$ is given by Eq.\ (\ref{eq:toy_de_identity_2}), its differential equation reads as
\begin{align}
 \partial_\ell f_0^{(2)}(\ell)&=-48f_2^{(1)}(\ell)f_2^{(1)}(\ell)=-48e^{-8\omega_0\ell}.
\end{align}
Using $f_0^{(2)}(0)=0$, it follows
\begin{align} f_0^{(2)}(\infty)=
-48\int\limits_0^{\infty}e^{-8\omega_0\ell}\mathsf{d}\ell=-\frac{6}{\omega_0}
\label{eq:toy_ep}.
\end{align}

In this example, we calculated and solved the perturbative flow equations separately in each order. For higher orders or more sophisticated systems, it is more advantageous to split the solution
into an algebraic task of deriving the DES and into a numerical task of solving it.
In the following, we discuss an efficient algorithm to handle the algebraic task for more general models and discuss its application to the uniform spin ladder.

\subsection{Generic algorithm}
\label{struct:algorithm}

\begin{figure}
 \includegraphics[width=\columnwidth]{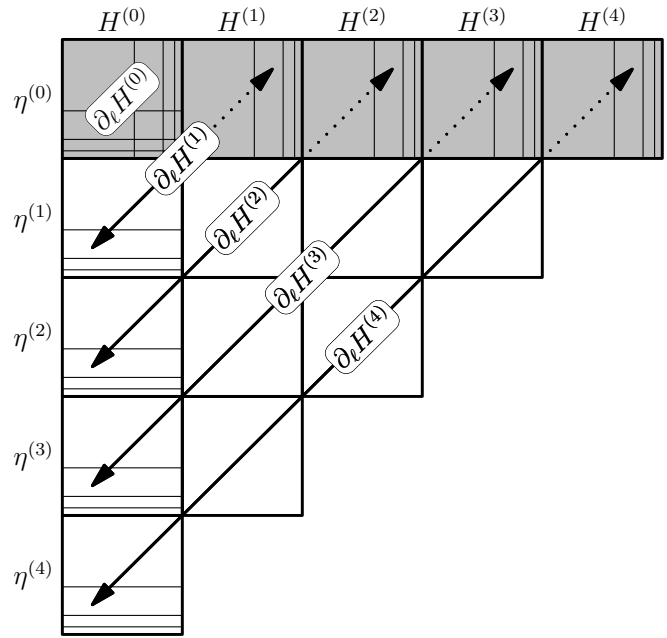}
 \caption{Sketch of the epCUT algorithm to calculate the DES for the iterative calculation of 
 $\partial_\ell H^{(4)}$. Due to the commutators 
 $[ \eta^{(1)},H^{(3)} ] , \dots , [ \eta^{(3)},H^{(1)} ] $, new terms with ${O_\text{min}^{}}=4$ emerge. 
 Thus, the calculation of the block $[ \eta^{(4)},H^{(0)} ] $ has to be carried out
 at last and self-consistently
 because it generates monomials contributing to the generator in the same order. If $H_0$ is not
  (block-)diagonal, both $[ \eta^{(4)},H^{(0)} ] $ and $[ \eta^{(0)},H^{(4)} ] $ have to be
  calculated simultaneously in a single self-consistent loop.}
 \label{img:algorithm_Hamiltonian}
\end{figure}

A key task in the implementation of epCUT is the design of an efficient algorithm to identify the monomials and
to calculate exactly the commutators which are relevant for the transformed Hamiltonian 
in the order of interest $n$. Henceforth, we call the order we are aiming at the
``targeted'' order.

Based on Eq.\ \eqref{eq:flow_perturbative}, we can calculate each order $m$ based on the results of lower orders.
Order zero is trivially given by the representation \eqref{eq:H-initial}
if $\widehat\eta [H_0]=0$, which means that $H_0$ is block diagonal.
The calculation of the commutators $\left[ \eta^{(1)},H^{(m-1)} \right] , \dots , \left[ \eta^{(m-1)},H^{(1)} \right] $ can be carried out independently, see Fig.\ \ref{img:algorithm_Hamiltonian}. 
According to Eq.\ \eqref{eq:flow_algeraic}, the commutator $\left[ \hat\eta[A_j],A_k \right] $ 
can be written as linear combination of monomials $A_i$ of which the prefactors define the contributions $D_{ijk}$ of the DES. 
For those monomials not yet present in the Hamiltonian, an additional 
monomial has to be included in the 
operator basis with a unique index.
We call the order in which a monomial occurs for the \emph{first} time its \emph{minimum} order ${O_\text{min}^{}}(A_i)$. 

We stress that in the evaluation of $\left[ \eta^{(p)},H^{(q)} \right] $, the commutator 
$\left[ \hat\eta[A_j],A_k \right] $ needs to be calculated only if ${O_\text{min}^{}}\left(A_i\right)=p$ and
${O_\text{min}^{}}\left(A_j\right)=q$.
For all monomials with lower ${O_\text{min}^{}}\left(A_i\right)$ and/or lower ${O_\text{min}^{}}\left(A_j\right)$, 
the commutators have already been calculated in lower orders.

The calculation of the commutators for $\left[ \eta^{(m)},H^{(0)} \right] $ is special because its result may include additional monomials of the same minimum order $m$ which were not considered so far. Since these monomials also enter the commutator via $\eta^{(m)}$, the block $\left[ \eta^{(m)},H^{(0)} \right] $ has to be iterated until no additional monomials occur: 
Then self-consistency is reached. This  should be done 
once the inner blocks $\left[ \eta^{(p>0)},H^{(q>0)} \right] $ are finished.

If the unperturbed Hamiltonian $H_0$ is local, the commutation of monomials from $\eta^{(m)}$ and $H_0$ lead to
monomials acting on the same local cluster or smaller subclusters. 
Furthermore, if the local Hilbert space of the cluster is finite, the number of new monomials which can be generated by iterative commutations with $H_0$ is bounded by the finite number of 
linearly independent matrices on this finite-dimensional Hilbert space.
Then the iterative loop is guaranteed to terminate after a finite number of cycles.

In the symmetric ladder model [see Eq. \eqref{eq:spinhammi}],
the local Hilbert spaces are finite so that a finite number of
cycles is sufficient. Even better, the commutation of the monomials
in terms of triplon creation and annihilation operators with
$H_0=J_{\bot}^{\text{e}}H_{\bot}^{\text{e}}+J_{\bot}^{\text{o}}H_{\bot}^{\text{o}}$
does not generate any additional monomials so that no iterations are needed in the calculation of
$\left[ \eta^{(p>0)},H^{(q>0)} \right]$.
These facts facilitate to reach high orders in the expansion parameter.

If the unperturbed Hamiltonian $H_0$ has also non-(block-)diagonal terms, 
the generator includes terms of order zero. Therefore, the blocks  $\left[ \eta^{(0)},H^{(m)} \right] $ have to be evaluated self-consistently as well. Since any term of the Hamiltonian may also appear in the generator, the
blocks $\left[ \eta^{(m)},H^{(0)} \right] $ and $\left[ \eta^{(0)},H^{(m)} \right] $ have to be calculated simultaneously within a joint self-consistency loop.
Self-consistency can be reached in a finite number of steps if the local
Hilbert space at each site is finite \emph{or} if the $H_0$ consists
of a sum of local bilinear bosonic or fermionic terms. Otherwise, it is difficult to 
see generally whether self-consistency can be reached.

For the sake of completeness, we note that in the special case $H_0:=H$, i.e., considering
the total Hamiltonian as the unperturbed one, the whole algorithm constructing the DES reduces to the calculation of the block $\left[ \eta^{(0)},H^{(0)} \right] $. This has to be done self-consistently with respect to both the generator and the Hamiltonian. 
This approach is the one employed in the self-similar CUT (sCUT) previously \cite{mielk97b,Fischer2010}. 
Since for $H_0=H$ ``the unperturbed'' part $H_0$
includes non-local and non-(block-)diagonal terms and perhaps refers even to an infinite local Hilbert space, the iteration of commutators will not terminate for any but the simplest models.
Thus, additional truncation criteria are needed, the validity of which needs to be justified.

\subsection{Perturbative evaluation of the uniform spin ladder}

\begin{table*}
 \begin{tabular}{|ccrclccc|}
\hline
$i$ 	&	 $A_i$						&	$h_i(0)$	&\hspace{0.8cm}	&	$h_i(\infty)$	&	${O_\text{min}^{}}$	&	${O_\text{max}^{0 \text{QP}}}$	&	${O_\text{max}^{1 \text{QP}}}$	\\				
\hline																							
0	&		$\sum\limits_r 	\mathbbm{1}			$	&	$-\nicefrac{3}{4}$	&	& $-\nicefrac{3}{4}-\nicefrac{3x^2}{8}$		&	0	&		2		&		2		\\
1	&		$\sum\limits_{r,\alpha} 	t^\dagger_{\alpha,r}t^{\phantom{\dagger}}_{\alpha,r}			$	&	$1$	&	& $1+\nicefrac{3x^2}{4}$		&	0	&		0		&		2		\\
2	&		$\sum\limits_{r,\alpha\neq\beta} 	t^\dagger_{\alpha,r}t^\dagger_{\alpha,r+1}t^{\phantom{\dagger}}_{\beta,r}t^{\phantom{\dagger}}_{\beta,r+1}			$	&	$-\nicefrac{x}{2}$	&	&	$-\nicefrac{x}{2}+\nicefrac{x^2}{8}$	&	1	&	\cellcolor[gray]{0.8} 	0		&	 \cellcolor[gray]{0.6} 	0		\\
3	&		$\sum\limits_{r,\alpha\neq\beta} 	t^\dagger_{\alpha,r}t^\dagger_{\beta,r+1}t^{\phantom{\dagger}}_{\beta,r}t^{\phantom{\dagger}}_{\alpha,r+1}			$	&	$\nicefrac{x}{2}$	&	&	$\nicefrac{x}{2}$	&	1	&	\cellcolor[gray]{0.8} 	-		&	 \cellcolor[gray]{0.6} 	-		\\
4	&		$\sum\limits_{r,\alpha} 	t^\dagger_{\alpha,r}t^\dagger_{\alpha,r+1}			 + \text{h.c.}$	&	$\nicefrac{x}{2}$	&	&	$0$	&	1	&		1		&		1		\\
5	&		$\sum\limits_{r,\alpha} 	t^\dagger_{\alpha,r}t^{\phantom{\dagger}}_{\alpha,r+1}			 + \text{h.c.}$	&	$\nicefrac{x}{2}$	&	&	$\nicefrac{x}{2}$	&	1	&	\cellcolor[gray]{0.8} 	-		&		2		\\
6	&		$\sum\limits_{r,\alpha\neq\beta} 	t^\dagger_{\alpha,r+2}t^{\phantom{\dagger}}_{\alpha,r}t^{\phantom{\dagger}}_{\beta,r+1}t^{\phantom{\dagger}}_{\beta,r+2}	+	t^\dagger_{\alpha,r}t^{\phantom{\dagger}}_{\beta,r}t^{\phantom{\dagger}}_{\beta,r+1}t^{\phantom{\dagger}}_{\alpha,r+2}	 + \text{h.c.}$	&	$0$	&	&	$0$	&	2	&	\cellcolor[gray]{0.8} 	-		&	 \cellcolor[gray]{0.6} 	-		\\
7	&		$\sum\limits_{r,\alpha\neq\beta} 	t^\dagger_{\beta,r+2}t^{\phantom{\dagger}}_{\alpha,r}t^{\phantom{\dagger}}_{\beta,r+1}t^{\phantom{\dagger}}_{\alpha,r+2}	+	t^\dagger_{\beta,r}t^{\phantom{\dagger}}_{\alpha,r}t^{\phantom{\dagger}}_{\beta,r+1}t^{\phantom{\dagger}}_{\alpha,r+2}	 + \text{h.c.}$	&	$0$	&	&	$0$	&	2	&	\cellcolor[gray]{0.8} 	-		&	 \cellcolor[gray]{0.6} 	-		\\
8	&		$\sum\limits_{r,\alpha} 	t^\dagger_{\alpha,r}t^{\phantom{\dagger}}_{\alpha,r+2}			 + \text{h.c.}$	&	$0$	&	&	$-\nicefrac{x^2}{8}$	&	2	&	\cellcolor[gray]{0.8} 	-		&		2		\\
9	&		$\sum\limits_{r,\alpha} 	t^\dagger_{\alpha,r+1}t^\dagger_{\alpha,r+2}t^{\phantom{\dagger}}_{\alpha,r}t^{\phantom{\dagger}}_{\alpha,r+1}		 + \text{h.c.}$	&	$0$	&	&	$\nicefrac{x^2}{4}$	&	2	&	\cellcolor[gray]{0.8} 	-		&	 \cellcolor[gray]{0.6} 	-		\\	
10	&		$\sum\limits_{r,\alpha\neq\beta} 	t^\dagger_{\beta,r+1}t^\dagger_{\alpha,r+2}t^{\phantom{\dagger}}_{\alpha,r}t^{\phantom{\dagger}}_{\beta,r+1}			 + \text{h.c.}$	&	$0$	&	&	$\nicefrac{x^2}{8}$	&	2	&	\cellcolor[gray]{0.8} 	-		&	 \cellcolor[gray]{0.6} 	-		\\
11	&		$\sum\limits_{r,\alpha} 	t^\dagger_{\alpha,r}t^\dagger_{\alpha,r+1}t^{\phantom{\dagger}}_{\alpha,r}t^{\phantom{\dagger}}_{\alpha,r+1}			$	&	$0$	&	&	$-\nicefrac{x^2}{4}$	&	2	&	\cellcolor[gray]{0.8} 	-		&	 \cellcolor[gray]{0.6} 	-		\\
12	&		$\sum\limits_{r,\alpha\neq\beta} 	t^\dagger_{\alpha,r}t^\dagger_{\beta,r+1}t^{\phantom{\dagger}}_{\alpha,r}t^{\phantom{\dagger}}_{\beta,r+1}			$	&	$0$	&	&	$-\nicefrac{3x^2}{8}$	&	2	&	\cellcolor[gray]{0.8} 	-		&	 \cellcolor[gray]{0.6} 	-		\\
13	&		$\sum\limits_{r,\alpha\neq\beta} 	t^\dagger_{\beta,r+1}t^\dagger_{\beta,r+2}t^{\phantom{\dagger}}_{\alpha,r}t^{\phantom{\dagger}}_{\alpha,r+1}			 + \text{h.c.}$	&	$0$	&	&	$\nicefrac{x^2}{8}$	&	2	&	\cellcolor[gray]{0.8} 	-		&	 \cellcolor[gray]{0.6} 	-		\\
14	&		$\sum\limits_{r,\alpha} 	t^\dagger_{\alpha,r+1}t^{\phantom{\dagger}}_{\alpha,r}t^{\phantom{\dagger}}_{\alpha,r+1}t^{\phantom{\dagger}}_{\alpha,r+2}			 + \text{h.c.}$	&	$0$	&	&	$0$	&	2	&	\cellcolor[gray]{0.8} 	-		&	 \cellcolor[gray]{0.6} 	-		\\
15	&		$\sum\limits_{r,\alpha} 	t^\dagger_{\alpha,r}t^\dagger_{\alpha,r+2}			 + \text{h.c.}$	&	$0$	&	&	$0$	&	2	&	\cellcolor[gray]{0.8} 	-		&	 \cellcolor[gray]{0.6} 	-		\\
16	&		$\sum\limits_{r,\alpha\neq\beta} 	t^\dagger_{\beta,r+1}t^{\phantom{\dagger}}_{\alpha,r}t^{\phantom{\dagger}}_{\beta,r+1}t^{\phantom{\dagger}}_{\alpha,r+2}			 + \text{h.c.}$	&	$0$	&	&	$0$	&	2	&	\cellcolor[gray]{0.8} 	-		&	 \cellcolor[gray]{0.6} 	-		\\
17	&		$\sum\limits_{r,\alpha\neq\beta} 	t^\dagger_{\alpha,r+1}t^{\phantom{\dagger}}_{\alpha,r}t^{\phantom{\dagger}}_{\beta,r+1}t^{\phantom{\dagger}}_{\beta,r+2}	+	t^\dagger_{\alpha,r+1}t^{\phantom{\dagger}}_{\beta,r}t^{\phantom{\dagger}}_{\beta,r+1}t^{\phantom{\dagger}}_{\alpha,r+2}	 + \text{h.c.}$	&	$0$	&	&	$0$	&	2	&	\cellcolor[gray]{0.8} 	-		&	 \cellcolor[gray]{0.6} 	-		\\
\hline
\end{tabular}
\caption{Basis operators $A_i$ (simple combinations of monomials obeying symmetry and/or hermiticity) occuring in
 a second order epCUT calculation for the uniform spin ladder using the particle conserving generator scheme ${\widehat \eta_\text{pc}}$. The third column contains the initial coefficients $h_i(\ell=0)$, the fourth the final renormalized coefficients $h_i(\ell=\infty)$. The minimum order ${O_\text{min}^{}}$ is given in which the corresponding operator 
 occurs for the first time; ${O_\text{max}^{0 \text{QP}}}$ is the highest relevant order of the coefficient
 for computing the ground-state energy and ${O_\text{max}^{1 \text{QP}}}$ is the highest relevant order
 for computing the dispersion (cf. Sec. \ref{struct:omax}).
 The terms marked in light gray are irrelevant for the computation of the
 ground-state energy in second order; the terms in dark gray are irrelevant if the dispersion is computed. If a term can not influence the targeted quantities at all, it has no maximal order (symbolized by a dash).
\label{tab:list_2nd_order}}
\end{table*}

\begin{table*}
  \begin{tabular}{rrrrrrrrrrrrrrrrrrrrrrrrr}
\hline
$i$	&		4		&		4		&	 \cellcolor[gray]{0.6} 	6		&	 \cellcolor[gray]{0.6} 	7		&		0		&		1		&	 \cellcolor[gray]{0.6} 	2		&	\cellcolor[gray]{0.8} 	8		&	 \cellcolor[gray]{0.6} 	9		&	 \cellcolor[gray]{0.6} 	10		&	 \cellcolor[gray]{0.6} 	11		&	 \cellcolor[gray]{0.6} 	12		&	 \cellcolor[gray]{0.6} 	13		&	 \cellcolor[gray]{0.6} 	14		&	 \cellcolor[gray]{0.6} 	15		&	 \cellcolor[gray]{0.6} 	16		&	 \cellcolor[gray]{0.6} 	17		&	 \cellcolor[gray]{0.6} 	6		&	 \cellcolor[gray]{0.6} 	7		&	 \cellcolor[gray]{0.6} 	14		&	 \cellcolor[gray]{0.6} 	15		&	 \cellcolor[gray]{0.6} 	16		&	 \cellcolor[gray]{0.6} 	17		&	\\	
$j$	&		4		&		4		&		4		&		4		&		4		&		4		&		4		&		4		&		4		&		4		&		4		&		4		&		4		&		4		&		4		&		4		&		4		&	 \cellcolor[gray]{0.6} 	6		&	 \cellcolor[gray]{0.6} 	7		&	 \cellcolor[gray]{0.6} 	14		&	 \cellcolor[gray]{0.6} 	15		&	 \cellcolor[gray]{0.6} 	16		&	 \cellcolor[gray]{0.6} 	17		&	\\	
$k$	&		1		&	 \cellcolor[gray]{0.6} 	2		&	 \cellcolor[gray]{0.6} 	2		&	 \cellcolor[gray]{0.6} 	3		&		4		&		4		&		4		&		4		&		4		&		4		&		4		&		4		&		4		&	\cellcolor[gray]{0.8} 	5		&	\cellcolor[gray]{0.8} 	5		&	\cellcolor[gray]{0.8} 	5		&	\cellcolor[gray]{0.8} 	5		&		1		&		1		&		1		&		1		&		1		&		1		&	\\	\hline
$D_{ijk}$	&		-2		&	 \cellcolor[gray]{0.6} 	-2		&	 \cellcolor[gray]{0.6} 	-1		&	 \cellcolor[gray]{0.6} 	-1		&		-6		&	\cellcolor[gray]{0.8} 	12		&	 \cellcolor[gray]{0.6} 	2		&	\cellcolor[gray]{0.8} 	-2		&	 \cellcolor[gray]{0.6} 	4		&	 \cellcolor[gray]{0.6} 	2		&	 \cellcolor[gray]{0.6} 	-6		&	 \cellcolor[gray]{0.6} 	2		&	 \cellcolor[gray]{0.6} 	-4		&	 \cellcolor[gray]{0.6} 	4		&	 \cellcolor[gray]{0.6} 	-2		&	 \cellcolor[gray]{0.6} 	2		&	 \cellcolor[gray]{0.6} 	1		&	 \cellcolor[gray]{0.6} 	-2		&	 \cellcolor[gray]{0.6} 	-2		&	 \cellcolor[gray]{0.6} 	-2		&	 \cellcolor[gray]{0.6} 	-2		&	 \cellcolor[gray]{0.6} 	-2		&	 \cellcolor[gray]{0.6} 	-2		&	\\	\hline
\end{tabular}
 \caption{Non-vanishing contributions $D_{ijk}$ to the differential equation system (DES) of the uniform spin
  ladder using the particle conserving generator scheme. 
  Operators and contributions marked in light gray are irrelevant for the computation of the
 ground-state energy in second order; those in dark gray are irrelevant for the dispersion. 
 The contributions are sorted by the commutators $[\hat\eta A_j,A_k]$ in which they are calculated.
   \label{tab:diffeq_2nd_order}}
\end{table*}

Here, we discuss the application of the generic algorithm to the uniform spin ladder [c.f. Eq.\ (\eqref{eq:spinhammi})] for a second-order calculation using the quasiparticle-conserving generator. \cite{knett00a,Fischer2010}

To evaluate the perturbation series for the ground-state energy or the dispersion relation of a sophisticated system, the first step is to write the Hamiltonian in second quantization and to identify the relevant monomials.
This operator basis $\left\{A_i\right\}$ is given in Table\ \ref{tab:list_2nd_order}
with $A_0$ and $A_1$ for the terms in $H_0$ (${O_\text{min}}= 0$) and $A_2$ to $A_5$ for the terms in 
$V$ (${O_\text{min}}=1$).
We combined certain monomials whose prefactors must be the same due to symmetry and/or hermiticity into one element of the operator basis $A_i$ (cf.\ Sec.\ \ref{struct:symmetries}). 
The advantage is that less operators need to be tracked.
The algorithm is not affected by this step except that the 
comparison of coefficients is a bit more complex.

Following the algorithm described above, the commutators of the block $\left[ \eta^{(1)},H^{(0)} \right] $ are calculated to complete the first order. The contributions to the DES obtained by comparison of coefficients are given in Tab.\ \ref{tab:diffeq_2nd_order}. Then, the contributions in second order are evaluated in the blocks $\left[ \eta^{(1)},H^{(1)} \right] $ and  $\left[ \eta^{(2)},H^{(0)} \right] $ leading to the new basis operators $A_{6-17}$ with ${O_\text{min}}=2$. 

Next, the perturbative flow equation \eqref{eq:flow_perturbative} has to be solved. We do this
numerically using a standard fourth order Runge-Kutta method \cite{NumRep}. The initial values for the coefficients in different orders of $x$ are read off the initial Hamiltonian. They are zero for all basis operators and all orders which are not present in the initial Hamiltonian.
We use a basis of only normal-ordered operators except for $A_0=\sum_r\mathbbm{1}$ so that the
 series expansion of the ground-state energy per rung $E_0$ is obtained in the limit of
 $\ell\to\infty$ from the prefactor of  $A_0$
\begin{subequations}
\begin{align}
 E_0 &=\sum\limits_{m=0}^{n} f_0^{m}(\infty)x^m+\mathcal{O}\left(x^{n+1}\right)
 \\
    &=-\frac{3}{4}-\frac{3}{8}x^2+\mathcal{O}\left(x^3\right).
\end{align}
\end{subequations}
Note that this result requires only three equations in the DES.

Likewise, the dispersion relation is determined from the renormalized coefficients of the hopping terms $A_1$,$A_5$ and $A_8$
\begin{subequations}
\begin{align}
 \omega(k)&= \sum\limits_{m=0}^{n}\Big( f_1^{m}(\infty)x^m+2f_5^{m}(\infty)x^m\cos(k)
\\
&\qquad +2f_8^{m}(\infty)x^m\cos(2k)\Big)+\mathcal{O}\left(x^{n+1}\right)\\
          &= 1+\frac{3}{4}x^2 + x\cos(k) -\frac{1}{4}x^2\cos(2k) +\mathcal{O}\left(x^3\right),
\end{align}
\end{subequations}
which require five equations, only two more than the ground-state energy.

\subsection{Optimizations}

\begin{figure*}
 \includegraphics[width=\columnwidth]{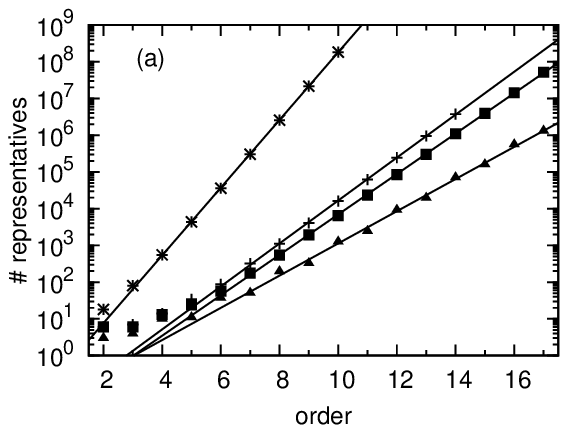}
 \includegraphics[width=\columnwidth]{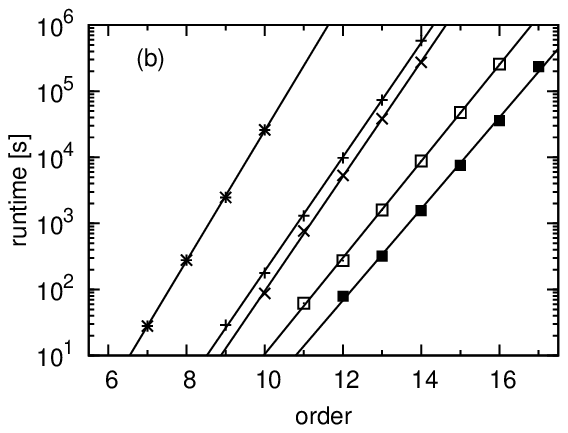}
  \caption{(a) Number of representatives in the effective Hamiltonian of the symmetric Heisenberg ladder vs.\ the order of
   the calculation for various optimizations aiming at the ground-state energy using ${\widehat \eta_\text{0}}$ 
   and all symmetries. Highest to lowest curve:
   full Hamiltonian, basic simplification rule, extended rule, full reduction of the DES based
   on the exact ${O_\text{max}^{}}$.
   (b) Runtime time for the construction of the DES vs.\ the order of the calculation with
   more and more optimizations using ${\widehat \eta_\text{0}}$ and all
   symmetries. Highest to lowest curve : full Hamiltonian without simplification, 
   basic \emph{a posteriori} simplification rule, extended \emph{a posteriori} rule, additional use of 
   the basic \emph{a priori} rule, additional use of the extended \emph{a priori} rule. The computations 
   were done on an Intel Xeon CPU (E5345, 2.33 GHz, single thread).}
  \label{plot:benchmarks}
\end{figure*}

\begin{table}
 \begin{tabular}{lc rrr}
 \hline
 generator & order & \# representatives & runtime & RAM\\
 scheme&&& [dd:hh:mm] & [GB] \\
 \hline
 0:n		&	17	&	 51,731,694	&	 2:17:14	&	8.1\\
 0:n, 1:n	&	15	&	107,513,297	&	13:09:12	&	17.3\\
 0:n, 1:n, 2:n	&	13	&	 51,371,642	&	11:09:47	&	8.0\\
 \hline
 \end{tabular}
 \caption{Number of representatives in the operator basis, total runtime and memory consumption for the symmetric Heisenberg ladder using various generator schemes in the highest order calculated. The computations were done on an Intel Xeon CPU (E5345, 2.33GHz, single thread) with full optimizations.}
 \label{tab:benchmark}
\end{table}

The epCUT method presented so far can be applied to a wide range of models in order to calculate a perturbative expansion of decoupled quasiparticle spaces. With increasing order, the number of representatives in the effective Hamiltonian, the runtime, and the memory consumption rise exponentially (see Fig.\ \ref{plot:benchmarks}). One is interested in increasing the order of
the calculation as high as possible because this generically enhances the accuracy of the
calculation: More and more orders kept imply that more and more physical processes with an increasing
spatial range are taken into account.

To increase the order, more efficient generator schemes and the symmetries known from sCUT can be exploited. Focussing on selected quantities of interest, the perturbative foundation of epCUT allows us to optimize the algorithm even further. Generic performance data possible with full optimizations are given in Table\ \ref{tab:benchmark}.

In practice, every optimization is carefully checked by comparing the results of the
optimized faster program to the results from the slower program 
before optimization. In this way, one can be sure that
no errors are introduced by incorrect assumptions.

\subsubsection{Generator scheme}

The (quasi)particle-conserving generator scheme  ${\widehat \eta_\text{pc}}$ used in our example decouples all
subspaces of differing numbers of excitations, i.e., quasiparticles, and sorts them in ascending order of their energy \cite{mielk98,uhrig98c,knett00a,Fischer2010}. In most applications,
however, only the ground-state and the low-lying excitations are of interest.
Consequently, the computational effort can be reduced by choosing a more efficient generator
scheme which targets the quantities of interest only. 
In 2010, Fischer, Duffe and Uhrig \cite{Fischer2010}
proposed a family of generator schemes based on modifications of ${\widehat \eta_\text{pc}}$ where 
only the first $q$ quasiparticle spaces are decoupled from the remaining
Hilbert space. The corresponding generator reads as
\begin{align}
 {\widehat \eta_{q}}\left[H(\ell)\right]:=
 \sum\limits_{j=0}^q \sum\limits_{i=j+1}^\infty \left( H^{i}_{j}(\ell)- H^{j}_{i}(\ell) \right).
\end{align}
In this notation, $H^{i}_{j}$ comprises all monomials of the Hamiltonian creating $i$ and annihilating $j$ quasiparticles.
For instance, the ground-state generator 
\begin{align}
 {\widehat \eta_\text{0}}\left[H(\ell)\right]=\sum\limits_i \left( H^{i}_{0}(\ell)- H^{0}_{i}(\ell)
 \right)
\end{align}
incorporates monomials which consist purely of either creation or annihilation operators.
Compared to the full quasiparticle-conserving generator, the effort to compute the 
corresponding DES is reduced significantly. 
In analogy to ${\widehat \eta_\text{pc}}$, the decoupled quasiparticle spaces are sorted according to energy.
Thus, the ground-state energy is given by the vacuum energy of the effective Hamiltonian
$H(\infty)$. If additionally the dispersion is calculated, the one-quasiparticle subspace has to be decoupled using ${\widehat \eta_{1}}$.
For decoupling higher quasiparticle spaces, analogous generator schemes can be used. But,
the increase in efficiency compared to the full quasiparticle-conserving generator becomes less
and less significant because the generator schemes $\widehat\eta_q$ do not conserve the
 block-band-diagonal structure of the Hamiltonian in contrast to ${\widehat \eta_\text{pc}}$.

\subsubsection{Symmetries}
\label{struct:symmetries}

For models defined on infinite lattices, it is necessary to use the translation symmetry
in order to be able to work directly in the thermodynamic limit.
In addition, the presence of other symmetries leads to linear dependencies of coefficients of
monomials which are linked by the symmetry transformations of the Hamiltonian. As in sCUT, \cite{reisc04,reisc06} this redundance can be significantly reduced by passing 
from simple monomials to symmetric linear combinations of them. 
Each of these polynomials is invariant under symmetry transformations of the Hamiltonian
and requires only one prefactor where the single monomials would need many more.
In our example (Table\ \ref{tab:list_2nd_order}) and in the following calculations
the size of the operator basis is reduced   by a factor of almost 24
 exploiting self-adjointness, reflection, and spin symmetry.

\subsubsection{Reduction of the differential equation system (DES)}
\label{struct:omax}

Targeting only certain quantities up to order $n$, such as the ground-state energy or the one-particle dispersion, the DES can be reduced. Here, we discuss how this can 
be done in practice.

Aside from the minimum order $O_{\min}$, a maximum order $O_{\max}$ can be a assigned to
each monomial and its coefficient $h_i$. The maximum order is the highest order of the series of
$h_i$ which still has an influence on the targeted coefficients up to order $n$. 
For instance, complicated processes involving many quasiparticles do not influence the ground-state energy directly, but only via other processes. Then, their $O_{\max}$ is much lower
than the targeted order $n$.
Technically, this is due to the hierarchy of the DES \eqref{eq:flow_perturbative}, which implies
\begin{subequations}
\begin{align}\label{eq:omax_des}
{O_\text{max}^{}}(A_j)&\geq {O_\text{max}^{}}(A_i)-{O_\text{min}^{}}(A_k)\text{,}\\
{O_\text{max}^{}}(A_k)&\geq {O_\text{max}^{}}(A_i)-{O_\text{min}^{}}(A_j)\text{,}
\end{align}
\end{subequations}
where the equality holds if we consider only a single contribution $D_{ijk}\neq 0$. The 
inequality takes into account that there may be many pairs $(i,k)$ for a given $j$. 
Thus, ${O_\text{max}^{}}(A_j)$ is the maximum value of all those right-hand sides:
\begin{align}
\label{eq:omax_it}
O_{\max}(A_j)=\max_{\{i,k|D_{ijk}\neq 0\}}\left[O_{\max}(A_i)-O_{\min}(A_k)\right] .
\end{align}
If $A_j$ is targeted,
for instance, the ground-state energy per rung $A_0$, its $O_{\max}$ is $n$ by definition.

For illustration, we consider the DES for the uniform spin ladder in second order (see Table\ \ref{tab:diffeq_2nd_order}). If we only target the ground-state energy $h_0$ up to order 2,
the maximum order of monomial $A_4$ is given by
\begin{subequations}
\begin{align}\label{eq:omax_condition}
{O_\text{max}^{}}(A_4)&= {O_\text{max}^{}}(A_0)-{O_\text{min}^{}}(A_4)= 2-1\\
\Rightarrow {O_\text{max}^{}}(A_4)&=1\text{,}
\end{align}
\end{subequations}
where we deal with equalities because there is only one contribution for $\partial_\ell h_0(\ell)$ in the DES and the ${O_\text{max}^{}}(A_0)$ is known. In this case, the maximum order $O_{\max}$ of $A_4$ is lower than the targeted order $2$.

The $O_{\max}$ of all coefficients can be calculated on the basis of the entire DES 
and of the minimum orders.
Note that Eq.\ \eqref{eq:omax_it} defines $O_{\max}$ implicitly, i.e., one has to 
find the correct self-consistent solution. This is done by starting from
\begin{align}
\label{eq:start}
O_{\max}(A_i)=
\begin{cases}
n, & \text{if $A_i$ is targeted} \\
0, & \text{otherwise.}
\end{cases}
\end{align}
A monomial $A_i$ is targeted if we want to compute its coefficient $h_i$ in the given order $n$. 
Starting from the initial choice \eqref{eq:start}, Eq.\ \eqref{eq:omax_it} is iterated:
the number $O_{\max}$ is increased if necessary until convergence is reached. 
Convergence is guaranteed because we consider a finite set of $\{A_i\}$  by construction and the $O_{\max}(A_i)$ are bounded from above by $n$. Hence, even in the worst case, there can be only a
finite number of increments. For illustration, 
the maximum orders for the uniform spin ladder in second order are given in Table\ \ref{tab:list_2nd_order} targeting dispersion or ground-state energy.

Once the maximum orders are known, we can reduce the DES because some coefficients have a maximum order lower than their minimum order:
\begin{align}\label{eq:reduce_1}
O_{\max}(A_i)<O_{\min}(A_i)\text{.}
\end{align}
Thus, they do not matter for the relevant quantities up to order $n$ and can be discarded completely. Moreover, all contributions to the DES which use these terms can be neglected. In addition, all contributions can be discarded for which 
\begin{align}
\label{eq:reduce_2}
O_{\max}(A_i)<O_{\min}(A_j)+O_{\min}(A_k)
\end{align}
holds. 

These considerations allow us to reduce the DES significantly. In Table\ \ref{tab:diffeq_2nd_order},
the reduction of the DES for the uniform spin ladder in second order is marked for 
the ground-state energy (light gray) and for the dispersion (dark gray), respectively.

We stress that one has to know the entire DES 
to apply the $O_{\max}$ concept as described above.

\subsubsection{Simplification rules}
\label{struct:basic_a-posteriori}

The reduction of the DES discards a large number of monomials and of the contributions $D_{ijk}$ [see for instance Fig.\ \ref{plot:benchmarks}(a)], which is essential for an efficient evaluation. 
But, it would be even more advantageous if one avoided the calculation of the omitted terms
\emph{before} they are tediously computed. The minimum orders ${O_\text{min}^{}}$ are known at each step of the iterative setup of the DES so that they can be used on the fly.
But, due to their implicit definition, the maximum orders ${O_\text{max}^{}}$ are not known
during the set-up of the DES.

Fortunately, estimates help. An upper bound for the maximum order is enough to accelerate the
algorithm, setting up the relevant part of the DES. Concomitantly, the memory consumption is reduced
significantly. \cite{hamer10} Henceforth, we call such estimates ``simplification rules''. Their
concrete form depends on the structure of the perturbed and the unperturbed Hamiltonian,
for instance, the block diagonality of the latter.  We emphasize that the simplification rules constitute
the part of the epCUT method which depends on the model.

In the following, we aim at a quantitative description up to order $n$ of the block of the effective
Hamiltonian pertaining to at most $q$ quasiparticles. 
For instance, $q=0$ provides the correct perturbative expansion of the ground-state energy and $q=1$ allows us to calculate the dispersion relation up to order $n$. 

A monomial creating $c$ triplons and annihilating $a$ triplons is targeted if both $c\le q$ and
$a\le q$ hold. Its maximum order is the targeted order ${O_\text{max}^{}}=n$. If it is not targeted, it can
influence the targeted terms by affecting terms consisting of fewer creation and annihilation operators via the DES. For the Heisenberg ladder, the unperturbed Hamiltonian [Eq. \eqref{eq:triplonhammi}] is block diagonal. Hence, no commutation of generator terms with $H_0$
changes  the number of created and annihilated triplons. The leading order of the generator
is 1, i.e.,  $\widehat\eta=\mathcal{O}\left(x\right)$.

In the commutation of a monomial with a generator term,
some of the local creation and annihilation operators may cancel due to normal ordering.
In order to yield a term affecting the first subspaces with $q$ 
quasiparticles,
\begin{subequations}
\begin{align}
 c^\prime = \max(c-q,0)
\end{align} 
local creation operators and  
\begin{align}
 a^\prime = \max(a-q,0)
\end{align}
\end{subequations} 
local annihilation operators have to cancel. 

First, we consider commutations with lowest-order generator terms stemming from the
initial Hamiltonian. In the spin ladder, these terms have order 1 and create or annihilate ${\Delta \text{QP}}=2$ quasiparticles on adjacent rungs. Because each commutation with $\eta^{(1)}$ 
increases the order of the affected coefficients by one, the maximum order is bounded by
\begin{align}
 {{\widetilde{O}}_{\max}} = n -\left\lceil {\frac{c^\prime}{2}} \right\rceil-\left\lceil {\frac{a^\prime}{2}} \right\rceil\ge {O_\text{max}^{}},
 \label{eq:basic_a-posteriori}
\end{align}
where the tilde on the left side means that one is dealing with an upper bound
and $\left\lceil {y} \right\rceil$ stands for the smallest integer that is still larger or equal to $y$.
If in the calculation of $\partial_\ell H^{(m)}$ the estimate ${{\widetilde{O}}_{\max}}$ of a monomial 
is lower than $m$, this contribution is irrelevant and can be omitted. This reduces the size of both the DES and of the Hamiltonian to be tracked. Moreover, discarding irrelevant monomials avoids the calculation of unnecessary commutators in the following iterations of the algorithm.

Clearly, the number of created and annihilated quasiparticles can be reduced by a number ${\Delta \text{QP}}$ larger than 2 by means of  commutations with generator terms involving more quasiparticles
which may have developed during the flow from the basic terms.
But, the generator terms involving more quasiparticles have a higher minimum order ${O_\text{min}^{}}$
so that a single commutation with them affects coefficients only in a higher order $m+{O_\text{min}^{}}$. 
In fact, for the used generator schemes, the ratio between ${\Delta \text{QP}}$ and ${O_\text{min}^{}}$ for new terms
developed during the flow can not exceed the corresponding ratio for generator terms present
in the initial Hamiltonian. Therefore, it is sufficient to consider only commutations with the initial terms in our simplification rules.

The above generic simplification rule can be easily adapted to other models as long as the unperturbed Hamiltonian $H_0$ is block diagonal. Otherwise, $H_0$ will lead to generator terms of order zero, which means that terms with high quasiparticle number can influence the coefficients of terms with low quasiparticle number in the same order. This is why it is desirable to
set up the perturbation in such a way that $H_0$ is block diagonal in the number of quasiparticles.

Applying the simplification rule reduces the number of representatives considerably (see Fig.\ \ref{plot:benchmarks}) leading to a significant improvement of runtime and memory consumption.
This basic simplification rule can be improved further by taking more model-specific information into account. A possibility to exploit the real-space structure of the monomials to lower the upper bound ${{\widetilde{O}}_{\max}}$ is described in Appendix \ref{struct:extended_a-posteriori}.

The computationally most costly part in the calculation of the DES is the evaluation of
commutators. Because the simplification rules sketched above can only be applied \emph{after} the commutation, we refer to them as \emph{a posteriori rules}. For the sake of efficiency, it is highly desirable to extend them to \emph{a priori rules}, estimating whether a commutator has to be evaluated at all \emph{prior} to its computation.
We describe such \emph{a priori} simplification rules in Appendixes \ref{struct:basic_a-priori} and \ref{struct:extended_a-priori}.

Because these rules are necessarily less strict than their \emph{a posteriori} analogs, one should use the combination of both kinds in practice.
The additional use of \emph{a priori} rules does not reduce the number of representatives or the memory consumption.  But, it boosts the speed of the calculation significantly because the vast
majority of commutators can be discarded, see [Fig.\ \ref{plot:benchmarks}(b)], and the a priori rules help to avoid the laborious computation of these unnecessary commutators.

\subsection{Directly evaluated epCUT}

\begin{figure}
 \includegraphics[width=\columnwidth]{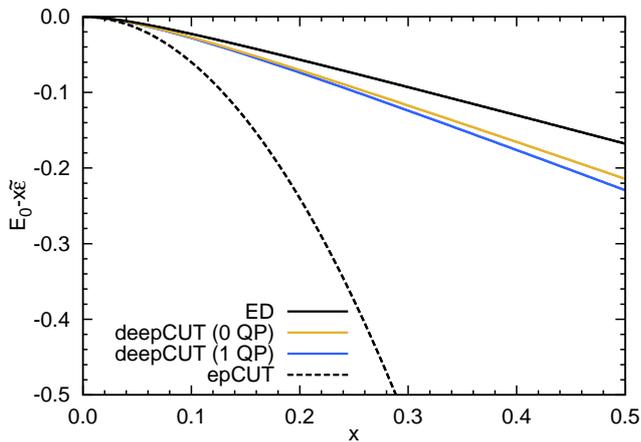}
 \caption{(Color online) Ground-state energy $E_0$ of the perturbed oscillator \eqref{eq:H_toy}
 relative to the first-order shift $x\tilde\epsilon$  vs.\ the expansion parameter $x$. For reference, the ground-state energy is also  determined by exact diagonalization (ED) considering 500 oscillator states (black solid line). The second-order result (black dashed line) deviates already significantly for small $x$ while the deepCUT results of the same order targeting the ground-state energy (0 QP) (light gray/orange line) and targeting additionally  the excitation energy (1 QP)(dark gray/blue line) are much more robust.
 The parameters are $\epsilon_0=0$, $\omega_0$=1, $\tilde\epsilon=10$, $\tilde\omega=12$, and $U=2$, (cf.\ 
 Sect.\ \ref{struct:model_toy}).}
\label{plot:E_0_toy}
\end{figure}

\begin{figure}
 \includegraphics[width=\columnwidth]{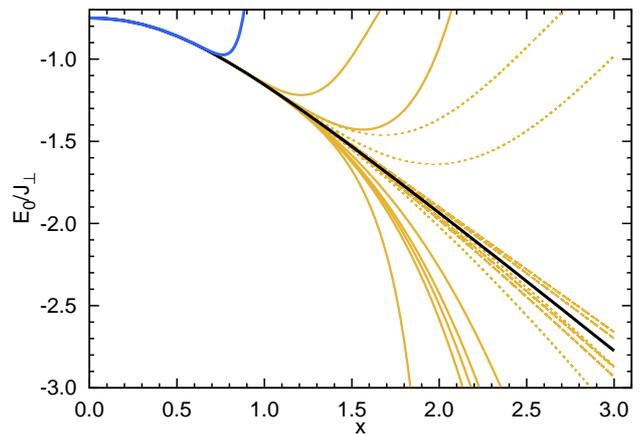}
 \caption{(Color online) 
 Ground-state energy per rung $E_0$ of the uniform spin ladder vs.\ relative leg coupling $x$ in order 17 using various evaluations. The direct evaluation (black line) renders a much more stable and reliable extrapolation of the plain perturbative series [dark gray (blue) line] than the various Pad\'e extrapolations [light gray (orange) line]. The solid light gray (orange) lines represent a standard  Pad\'e extrapolation, the dotted line a Pad\'e extrapolation  in $u$ ($x=\nicefrac{u}{1-u}$), and the dashed line a Pad\'e extrapolation in $u$ including the asymptotic behavior of the spin ladder given by the ground-state energy of the spin chain.}
 \label{plot:E_0_perturbative_pade_direct}
\end{figure}

\begin{figure}
 \includegraphics[width=\columnwidth]{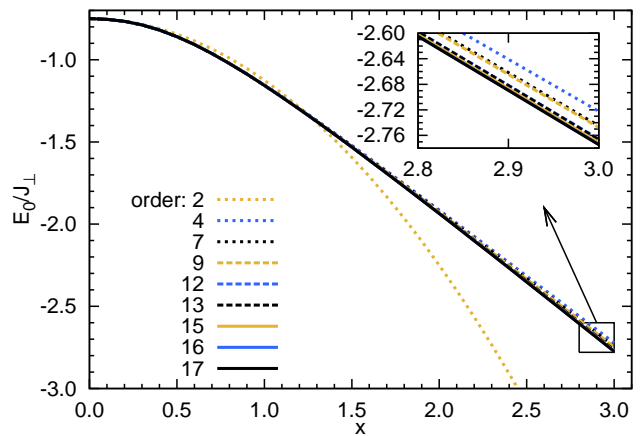}
 \caption{(Color online) Ground-state energy per rung $E_0$ of the uniform spin ladder vs.\ 
  relative leg coupling $x$ for different orders using the direct evaluation.}
 \label{plot:E_0_direct_orders}
\end{figure}

\begin{figure}
 \includegraphics[width=\columnwidth]{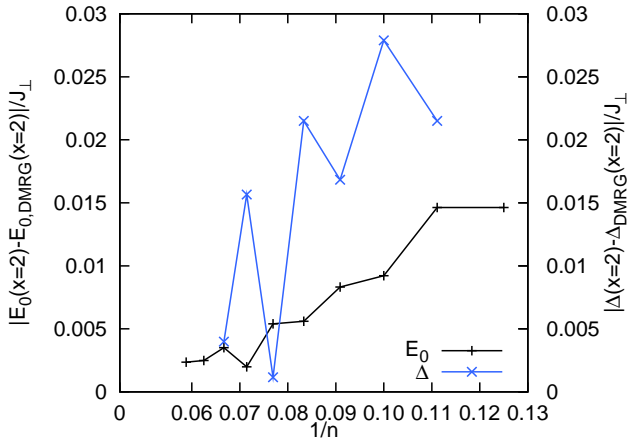}
 \caption{(Color online) Deviations between the results (ground-state energy and spin gap) of the deepCUT and of a DMRG (density matrix renormalization group) 
 calculation for the uniform spin ladder vs.\  the inverse order $1/n$ for $x=2$. }
 \label{plot:1_over_n}
\end{figure}

In addition to the perturbative evaluation, the reduced DES computed by epCUT 
in a given order $n$ can be evaluated non-perturbatively.
After the reduction step described in Sec.\ \ref{struct:omax}, the DES consists exclusively of contributions which are relevant to the targeted quantities in the desired order $n$. 
This reduced DES in Eq.\ \eqref{eq:flow_numeric} can be numerically integrated for any given value of $x$ to obtain the coefficients of the Hamiltonian $h_i(\ell)$ \emph{directly} without passing by an expansion in $x$. In such a calculation, all coefficients influence one another to infinite order.
The numerical solution depends on the expansion parameter in an intricate manner and can no longer be understood as \emph{finite} partial sum of an infinite series.
In this sense, the perturbative reduced DES in order $n$ is extrapolated by the
direct evaluation in a non-perturbative way. To stress the difference 
to perturbation series computed by  epCUT, we call this technique 
\emph{directly evaluated epCUT} (deepCUT). We keep the term ``enhanced perturbative''
in this expression because the approach is derived from the epCUT, and the perturbative order
of the epCUT determines the spatial range of physical processes captured. Yet, we stress
that by the direct evaluation contributions to infinite order in $x$ are included.

We emphasize that the reduction of the DES \emph{before} the numeric integration is essential.
It enhances the performance of the integration because the reduced DES is much smaller.
But, the crucial observation is that the reduction renders the integration much more robust.
Numerical integrations of the full DES diverge for high orders and high values of $x$. 
We conclude that the reduced DES represents the relevant 
physical processes in a more consistent way. The integration of the full DES 
generates spurious higher-order contributions which overestimate certain effects. In an exact
solution, the spurious higher-order contributions would be compensated by other 
processes which are captured only in a higher-order calculation. 

Analogous observations are
known from diagrammatic perturbation theory where the inclusion of subsets of
diagrams in infinite order does not guarantee improved results. Improved results can only be
expected from systematically controlled calculations. The inclusion of infinite orders is indicated
if this achieves conserving self-consistent approximations. For instance,
the shift of poles in a propagator is not captured by any finite perturbation series
in the propagator, but it follows easily from a perturbation of the self-energy. \cite{ricka80}

We show the difference between deepCUT and epCUT for the perturbed harmonic oscillator \eqref{eq:H_toy}. 
Targeting the ground-state energy, the first step is to calculate the maximal orders of the representatives
$A_i$  and to reduce the contributions in the DES to the relevant ones (cf.\ Tables\ \ref{tab:toy_list_2nd_order} and \ref{tab:toy_diffeq_2nd_order}). 
The  minimal DES for the coefficients $h_i$ of the three relevant representatives in second order reads as
\begin{subequations}
\label{eq:toy_DES}
\begin{align}
 \partial_\ell h_0 &= -48 h_2 h_2,		&	h_0(0)&=\epsilon_0 + \tilde\epsilon x	\\
 \partial_\ell h_1 &= 0,			&	h_1(0)&=\omega_0 + \tilde\omega x&=h_1(\ell)	\\
 \partial_\ell h_2 &= -\phantom{0}4 h_2 h_1,	&	h_2(0)&=x.
\end{align}
\end{subequations}
In contrast to the epCUT, different powers of the expansion parameter $x$ are not split.
Because $h_1(\ell)$ remains constant, the coefficient in the generator
 can be determined analytically as
\begin{align}
 h_2(\ell)=h_2(0)e^{-4h_1(0)}=xe^{-4(\omega_0+6\tilde\omega x)}.
\end{align}
For the ground-state energy, it follows that
\begin{subequations}
\label{eq:toy_deep_qp0}
\begin{align}
  h_0(\infty)&=h_0(0)-48x^2\int\limits_0^{\infty}e^{-8h_1(0)}\mathsf{d}\ell\\
 &=\tilde\epsilon x-\frac{6x^2}{\omega_0+x\tilde\omega}.
\end{align}
\end{subequations}
At first glance, the slight modification $\omega_0\to\omega_0+x\tilde\omega$ in the energy denominator compared to the perturbative second order result \eqref{eq:toy_ep} seems inconspicuous. But, we stress that a Taylor series of \eqref{eq:toy_deep_qp0} includes infinite orders of $x$. In Fig.\ \ref{plot:E_0_toy}, the results are compared to exact diagonalization (ED) in the Hilbert space of 500 states. Even for small values of the expansion parameter, the perturbative result deviates significantly while the deepCUT of the same order behaves reasonable even at $x=0.5$ and beyond. We stress that the fact that we can
solve the equations analytically is due to the simplicity of the calculations for
this particular model in low order.

The perturbative result for the ground-state energy $h_0$ does not depend on whether or not
we target on the single excitation energy $h_1$. This is different in deepCUT where changes
in the DES due to varying targeted quantities will generally influence all quantities, at least
weakly. Targeting both the ground-state energy $h_0$ and the excitation energy $h_1$ modifies the 
derivative of $h_1(\ell)$ to 
\begin{align}
\label{eq:toy_flow_qp1}
 \partial_\ell h_1 &= -192 h_2 h_2,			&	h_1(0)&=\omega_0 + \tilde\omega x	
\end{align}
so that now the complete DES is given by \eqref{eq:toy_DES} and by \eqref{eq:toy_flow_qp1}.
We solve the DES similar to a previous treatment \cite{dusue04a} introducing the  quantity 
\begin{align}
 \Omega=\sqrt{h_1^2-48h_2^2}
\end{align}
which is conserved along the flow. Physically meaningful values are $\Omega^2\ge 0$.
Both $h_1$ and $h_2$ decrease during the flow until $h_2$ vanishes in the limit of infinite $\ell$. Then the effective Hamiltonian reads as
\begin{subequations}
 \begin{align}
  h_0(\infty)&={\Omega},\\
  h_1(\infty)&=\frac{1}{4}\left({\Omega}-\omega_0-\tilde\omega x\right)+\epsilon_0+\tilde\epsilon x,\\
  h_2(\infty)&=0.
 \end{align}
\end{subequations}
As can be seen in Fig.\ \ref{plot:E_0_toy}, targeting $h_1$ as well modified the result for $h_0$, although only slightly.

Next, we illustrate the deepCUT for the extended model of the uniform spin ladder.
Figure \ref{plot:E_0_perturbative_pade_direct} compares the ground-state energy per rung
$E_0$ of the uniform spin ladder as function of the relative leg coupling $x$ obtained from the 
plain perturbative  series in order 17, from various Pad\'{e} extrapolations, and from
the direct evaluation. 

We use three different kinds of Pad\'e extrapolations: First, a standard Pad\'{e} extrapolation for the series expansion of the ground-state energy in $x$; second, an extrapolation for $(1-u)E_{0}(u)$, where we rewrite the expansion paramater as x=\nicefrac{u}{(1-u)}; third,
a Pad\'{e} extrapolation for $(1-u)E_{0}(u)$ including the asymptotic  behavior of the spin ladder. For $x\to\infty$ one obtains two isolated spin chains whose ground-state energy per site $e_0=\nicefrac{1}{4}-\ln 2$ is known.
\cite{Bethe1931,Hulthen1938} Thus $E_0(x)\to 2e_0 x + {\cal O}(x^0)$ for $x\to\infty$.

The plain series shoots up at about $x\approx 0.7$ while the Pad\'{e} extrapolations start to scatter strongly beyond $x\approx 1$.
The direct evaluation lies between the two stiffest Pad\'{e} extrapolations and remains stable 
up to even very large values of the expansion parameter $x\approx3$. Comparing various orders,
see Fig.\ \ref{plot:E_0_direct_orders}, the results of deepCUT converge rapidly and display only minor corrections for large values of $x$ indicating a high reliability. 
The convergence of the ground-state energy and the spin gap with increasing order $n$ is
 displayed in Fig.\ \ref{plot:1_over_n}. Clearly, increasing the order improves the results,
 but the convergence is not monotonic. In the spin gap an even-odd effect is visible.
In Sec.\ \ref{struct:results_uniform}, further comparisons of the deepCUT results with those of other methods will be presented.

This deepCUT bears similarities to the sCUT approach
\cite{mielk97b,reisc04,reisc06,Fischer2010,dresc11,Duffe11}.
In sCUT, a set of basis operators is selected by a \emph{truncation scheme} and for
this set the full DES is computed. It comprises all commutation relations between the
selected basis operators.
In deepCUT, the order of the expansion parameter takes over the role of the truncation scheme.
But we stress that deepCUT is not self-similar: In sCUT all commutators between the selected monomials are considered. In epCUT and thus in deepCUT only the commutators between specific subblocks based on the minimum orders ${O_\text{min}^{}}$ are considered, see Sec.\ \ref{struct:algorithm}.
Moreover, targeting certain subspaces with $q$ quasiparticles and the concomitant
reduction of the DES does not only discard irrelevant monomials. Also contributions linking relevant monomials are canceled if their effect is of too high order.
Therefore,  the \grq truncation\grq\ taking place in (de)epCUT, controlled by the expansion parameter, is a \emph{truncation of the DES} rather than a \emph{truncation of operators} as it is done in the sCUT approach.

One practical advantage of the deepCUT over the sCUT is that only one parameter, the maximum order
of the expansion parameter, needs to be fixed in order to define the approximation.
In the sCUT, generically many parameters define the truncation scheme \cite{reisc04,reisc06,Fischer2010,dresc11,Duffe11} which leaves some ambiguity about how to
systematically improve the approximation.

Another comparison of approaches is in order. Recently, Yang and Schmidt proposed a CUT approach 
based on graph theory (gCUT). \cite{yang11a} Their approach generalizes an idea 
first put forward by Irving and Hamer for static ground-state properties under the name
of ``exact linked cluster expansion'' (ELCE). \cite{irvin84}
Yang and Schmidt are able to treat effective models quite generally.
To compute a certain quantity such as the ground-state
energy the irreducible contributions of subgraphs, i.e., of linked clusters,
of the lattice are summed. The size of the largest subgraph considered determines the
approximation. The larger it is the better the system is described because physical processes
with a larger range are kept. Thus the fundamental idea of the approach is similar to the
one of deepCUT: Truncation in the range of processes, but local processes are kept to
infinite order. 

The main difference is that the actual CUT is done on clusters.
So ELCE and gCUT have advantages and disadvantages. An advantage is that it is sufficient
to deal with finite dimensional Hilbert spaces and the transformations can be performed on 
matrices. A disadvantage is that momentum conservation cannot be exploited on the
level of the clusters because they are finite which restricts the choice of applicable
generators. \cite{yang11a}
The deepCUT is based on second quantization \cite{knett03a} and can take advantage of all
symmetries of the problem under study.
A detailed comparison of the approaches is left to future studies.

\subsection{Transformation of Observables}
\label{struct:observables}

\begin{figure}
 \includegraphics[width=\columnwidth]{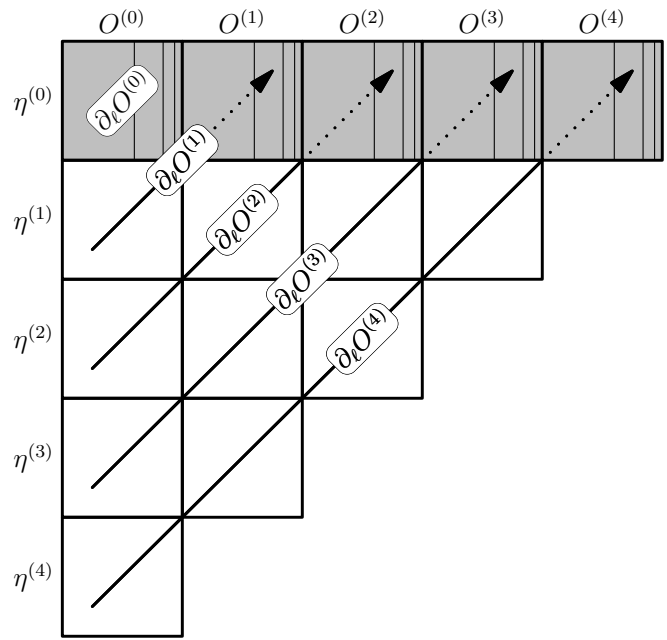}
 \caption{Sketch of the epCUT algorithm to calculate the DES for $\partial_\ell O^{(4)}$ iteratively. Due to the commutators $[ \eta^{(1)},O^{(3)} ] , \dots , [ \eta^{(4)},O^{(0)} ] $, additional
  terms with ${O_\text{min}^{}}=4$ emerge. In contrast to the algorithm for the Hamiltonian, see Fig.\ \ref{img:algorithm_Hamiltonian}, no self-consistent calculation is
needed for $[ \eta^{(4)},O^{(0)} ] $. Self-consistency is required only for $[ \eta^{(0)},O^{(4)} ] $ if $\eta^{(0)}$ is finite.}
 \label{img:algorithm_observable}
\end{figure}

In order to calculate spectral densities for instance, the coefficients of the corresponding
 observable must be known with respect to the same basis as the effective Hamiltonian. Thus the observables must be transformed as well. This can be realized by integration of the 
flow equation for observables
\begin{align}
 \partial_\ell O(\ell) = \left[ \eta(\ell),O(\ell) \right] 
 \label{eq:flow_obs}
\end{align}
introduced by Kehrein and Mielke \cite{kehre97,kehre98a}.

In analogy to the transformation of the Hamiltonian discussed in Sec.\ \ref{struct:flow_Hamiltonian}, we introduce an operator basis $B_i$ for the observable shifting the dependence on $\ell$ from the operators to their coefficients
\begin{subequations}
\begin{align}
 O(\ell) &=\sum\limits_i o_i(\ell)B_i\\
  &= \sum\limits_{i}\sum\limits_{m=0}^n f_i^{(m),\text{obs}}(\ell)x^m B_i,
  \label{eq:obs_series}
\end{align}
\end{subequations}
where the second equation stands for the perturbative expansion of these coefficients.
Hence the flow equation for observables \eqref{eq:flow_obs} leads to a DES for their
coefficients
\begin{subequations}
\begin{align}
 \partial_\ell o_i(\ell) = \sum\limits_{j,k} D_{ijk}^{\text{obs}}h_j(\ell)o_k(\ell).
\label{eq:flow_obs_numeric}
\end{align}
The contributions $D_{ikj}^{\text{obs}}$ are obtained by calculating the commutators between the monomials of the generator and the monomials of the observable followed by a comparison of the coefficients
\begin{align}
 \sum\limits_i D_{ikj}^{\text{obs}} B_i = \left[ \hat\eta [A_j],B_k \right] .
\label{eq:flow_obs_algeraic}
\end{align}
\end{subequations}

The differential equations \eqref{eq:flow_obs_numeric} imply a hierarchical DES for the perturbative series \eqref{eq:obs_series} for the coefficients
\begin{align}
 \partial_\ell f_i^{(m),\text{obs}}(\ell) = \sum\limits_{jk}\sum\limits_{p+q=m} D_{ijk}^{\text{obs}} f_j^{(p)}(\ell) f_k^{(q),\text{obs}}(\ell).
\label{eq:flow_obs_perturbative}
\end{align}

The algorithm for the calculation of the DES in Sec.\ \ref{struct:algorithm} can easily be
adapted for the transformation of observables. 
Each order of the differential $\partial_\ell O^{(m)}$ is calculated recursively,
cf.\ Fig.\ \ref{img:algorithm_observable}.
Since the generator  $\eta$ is defined solely by the Hamiltonian, it is not influenced by the
outcome of the transformation of observables.
For this reason the evaluation of $\left[ \eta^{(m)},O^{(0)} \right] $ 
does not need to be carried out self-consistently. 
After the calculation of the commutators 
$\left[ \eta^{(1)},O^{(m-1)} \right] \dots\left[ \eta^{(m-1)},O^{(1)} \right] $, only the block $\left[ \eta^{(0)},O^{(m)} \right] $ has to be treated self-consistently.
But recall that $\eta^{(0)}$ only occurs if the unperturbed 
Hamiltonian $H_0$ is not (block-)diagonal.
Because both differential equations \eqref{eq:flow} and \eqref{eq:flow_obs} are coupled
by the generator, their integrations have to be done simultaneously.

For the transformation of the Hamiltonian, we extensively discussed that only
certain contributions really matter. We introduced the concept of a maximum order
in which the coefficient of a physical process needs to be known in order to
influence the targeted quantities. This concept allowed us to reach 
significantly higher orders. Thus, we want to extend the concept of a maximum
order also to the transformation of observables.
It turns out that this extension is rather subtle.

Before, in the flow of the Hamiltonian, the maximum order of a generator coefficient
 ${O_\text{max}^{\eta,H}}(A_i)$ is the maximum order of the same monomial ${O_\text{max}^{H}}(A_i)$ in the Hamiltonian. Now, we also target certain blocks of the observable and they
 are influenced by the monomials in the generator. This leads to maximum orders for both the observable term ${O_\text{max}^{O}}(B_i)$ and the generator terms ${O_\text{max}^{\eta,O}}(A_i)$. 
 The latter does not need to coincide with the maximum order  ${O_\text{max}^{\eta,H}}(A_i)$ resulting
 from the consideration of the Hamiltonian flow alone.
Thus, one has to find a unique and unambiguous way to fix ${O_\text{max}^{\eta}}(A_i)$. 
We discuss three alternatives:

(A) The maximum order of the generator terms is chosen in such a way that the targeted quantities in
 both the Hamiltonian and the observable(s) can be computed up to the targeted order
 \footnote{One may also target at order $n_H$ for the Hamiltonian and a different
 order $n_O$ for the observable.}  $n$
\begin{align}
 {O_\text{max}^{\eta}}(A_i)=\max ({O_\text{max}^{H}}(A_i),{O_\text{max}^{\eta,O}}(A_i)).
\end{align}
Then the iterative calculation of the ${O_\text{max}^{}}$ must be realized within a single self-consistent
loop. The perturbative evaluation yields a perturbative series for the coefficients of the observables under the transformation with the \emph{full} generator up to order $n$. 
It may happen that in this way some generator terms are assigned a higher
${O_\text{max}^{\eta}}\ge{O_\text{max}^{\eta,H}}$ than in the transformation of the Hamiltonian alone so that 
the DES of the Hamiltonian comprises additional contributions. 
By construction, this does not affect the perturbative evaluation of the epCUT.
But it will affect its direct evaluation (deepCUT) although it should be absolutely
minor in a parameter regime of good convergence of the flow.

(B) Alternatively, the determination of ${O_\text{max}^{O}}(B_i)$ and ${O_\text{max}^{\eta,O}}(A_i)$ can be realized after and strictly separated from the calculation of ${O_\text{max}^{\eta,H}}(A_i)$ and ${O_\text{max}^{H}}(A_i)$. Monomials which are discarded due to the reduction of the Hamiltonian will not be considered for the DES of the observables even though 
this may affect the targeted coefficients of the observable.
Hence the transformation of the observables in perturbative evaluation is not realized with respect to the complete generator. 
We stress that this does not violate the unitarity of the transformation up to the calculated order 
because the generator is still anti-Hermitian and it is essentially 
the same as for the transformation of the Hamiltonian. 
No significant deviations are expected in the regime of good convergence of the
flow.
Note also that any generator whose coefficients differ only by orders larger than ${O_\text{max}^{H}}(A_i)$ 
leads to the same perturbative series for the relevant quantities in the Hamiltonian.

(C) A third alternative consists in taking over the ${O_\text{max}^{\eta,H}}(A_i)$ for the 
reduction of the DES for the observables. Then only the values ${O_\text{max}^{O}}(B_i)$ are computed self-consistently.

For deepCUT, alternatives (B) and (C) ensure that the DES for the Hamiltonian 
is independent of the considered observables. Generally, we expect that the precision
in the derivation of effective Hamiltonians is more important than the precision
of matrix elements. Also in experiment, energies are generically known to much higher
accuracy than matrix elements.

In order to keep the effective Hamiltonian in direct evaluation independent of the observables, we decide to use alternative (B) for deepCUT.
For the perturbative evaluation, however, we favor alternative (A) because it makes
the rigorous determination of the perturbation series of matrix elements possible.

The computational performance can again be increased decisively by applying simplification rules. 
They can be used directly for observables if both the Hamiltonian and the observables meet their requirements. This can often be achieved by appropriate definitions.
For instance, the observable 
\begin{align}
2S_0^{\text{L},z}=t^\dagger_{z,0}+t^{\phantom{\dagger}}_{z,0}+it^\dagger_{y,0}t^{\phantom{\dagger}}_{x,0}-it^\dagger_{x,0}t^{\phantom{\dagger}}_{y,0}
\end{align}
is needed for the calculation of the dynamic structure factor relevant for inelastic
neutron scattering. But this observable includes non-block-diagonal terms in order zero. 
To circumvent this problem, we consider the observable $x\cdot S_0^{\text{L},z}$ instead. 
In this way, the non-block-diagonal monomials in the observable are shifted to order 1
so that they behave like the non-block-diagonal perturbation in the Hamiltonian.
One loses an order of accuracy for a given fixed order $n$ of the calculation. 
But, all the simplification rules relying on block diagonality in order zero can be used
as before.

\section{Results for uniform spin ladder}
\label{struct:results_uniform}

\subsection{Ground-state energy}
\begin{figure}
 \includegraphics[width=\columnwidth]{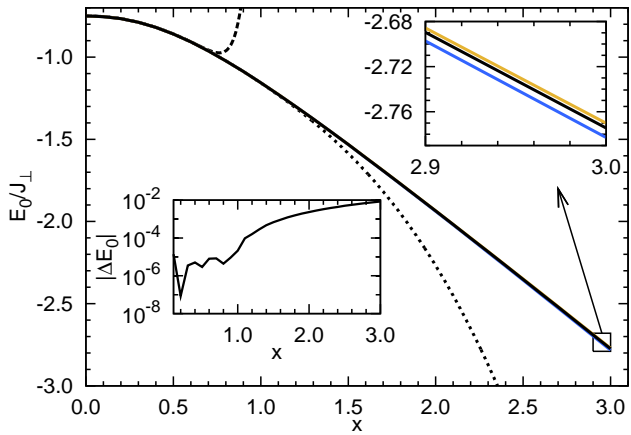}
 \caption{(Color online) Ground-state energy $E_0$ per rung of the uniform spin ladder
  vs.\ relative leg coupling $x$ resulting from
  various methods. The epCUT results (order 17; direct: black, solid; perturbative: 
  black, dashed; Pad\'{e}[11,6]: black, dotted) agree with the sCUT results
 ($d$=(12,10,10,6,6,5,5,4,4), light gray (orange), solid) and the DMRG results 
 (dark gray (blue)). The energies from the direct evaluation, the sCUT and the DMRG lie on top
 of each other, see also upper inset. The deviation $|\Delta E_0| =
  |E_{0,\text{direct}}-E_{0,\text{DMRG}}|$ is shown in the lower inset.}
 \label{plot:E_0_ladder_com}
\end{figure}

The ground-state energy per rung is calculated up to and including order 17. 
The results of the direct (black solid line) and of the perturbative evaluation 
(dashed black line) are displayed in Fig.\ \ref{plot:E_0_ladder_com}. The coefficients of the perturbative series, see Tab.\ \ref{tab:coefficients_gse}, agree perfectly with the fractions from
pCUT (up to order 14) \cite{knett03b} and with the decimal numbers (up to order 23) 
given by Zheng et al.\ \cite{Zheng1998}.
This agreement shows that the epCUT works for a system with equidistant spectrum in $H_0$.

The plain series is trustworthy only up to $x\approx0.7$. For larger $x$, extrapolations are needed.
The dotted black line shows the ``best'', i.e., stiffest,  Pad\'{e} extrapolant of order $[11,6]$ 
for this series. Other Pad\'{e} extrapolants are shown in Fig.\ \ref{plot:E_0_perturbative_pade_direct}.
The results of the deepCUT are depicted as solid black line in Fig.\ \ref{plot:E_0_ladder_com}. 
It fits perfectly to the perturbative result for weak leg couplings. For larger values of $x$,
it serves as an excellent extrapolation of the perturbative results. In order to support
that the directly evaluated  results are quantitatively reliable even for larger $x$, the deepCUT data are compared with results from sCUT (Refs.\onlinecite{Fischer2010,Duffe10}) and from DMRG. \cite{Raas2011,ALPS_1,ALPS_2} The sCUT results, represented by the solid light gray (orange)
line, are calculated with the ground-state generator and the truncation scheme $d$=(12,10,10,6,6,5,5,4,4). The truncation reads $d=(d_2,d_3,d_4,\dots)$, where $d_i$ denotes the
real space extension of a monomial with $i$ interacting quasiparticles. A monomial with $i$ interacting quasiparticles is truncated if it exceeds the extension $d_i$. For instance, the
monomial $t^\dagger_{\alpha,r}t^{\phantom{\dagger}}_{\alpha,r+4}$ has an extension $d_2=4$.

The DMRG data, represented by a solid dark gray (blue) line, results from a finite-size scaling.
The ground-state energies for ladders with $L=40,60,\dots,160$ rungs and $m=500$ states are extrapolated with the ansatz
\begin{align}
\label{eq:ansatz_DMRG}
E_0(L)=E_0(\infty)+c_0\frac{e^{-L/L_0}}{L^{c_1}}
\end{align}
to estimate the ground-state energy for an infinity ladder $E_{0}(\infty)$ for each value of $x$.
 
The results of these methods agree perfectly. The deviations between the DMRG results and the results of the direct evaluation are shown in the lower inset of Fig.\ \ref{plot:E_0_ladder_com}. They increase with rising $x$, but they remain still small. For $x=1.5$, the deviation is less than $10^{-3}J_\perp$ and for $x=3$ it is still less than $10^{-2}J_\perp$.

\subsection{Dispersion}

\begin{figure}
 \includegraphics[width=\columnwidth]{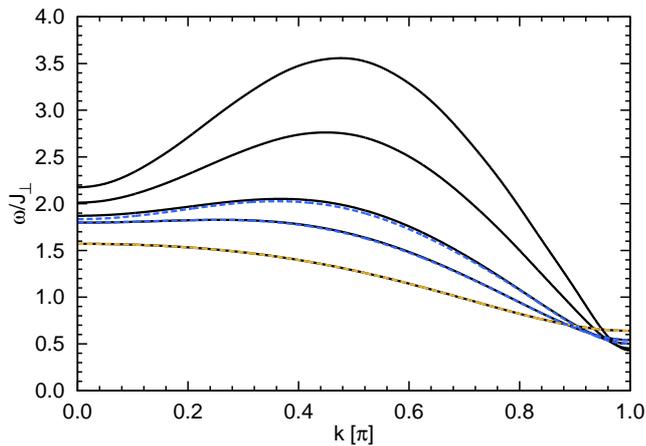}
 \caption{(Color online) Dispersion $\omega(k)$ of the uniform spin ladder for various values of $x\in\{0.5,0.8,1,1.5,2\}$ and various evaluation techniques based on order 15. At
$k=\nicefrac{\pi}{2}$, the lowest curve is $x=0.5$ and the highest curve is $x=2$. For $x=0.5$,
the plain series in $x$ (light gray (orange), dashed) and for $x=0.8$ and $x=1$ the plain series in the  parameter $p(x)$ from Eq.\ \eqref{eq:internal_parameter} (dark gray (blue),
dashed) are shown. The results of the direct evaluation (black, solid) agree well  with the perturbative results for small $x$ and they are still robust for larger $x$.}
 \label{plot:Dispersion_ladder}
\end{figure}

\begin{figure}
 \includegraphics[width=\columnwidth]{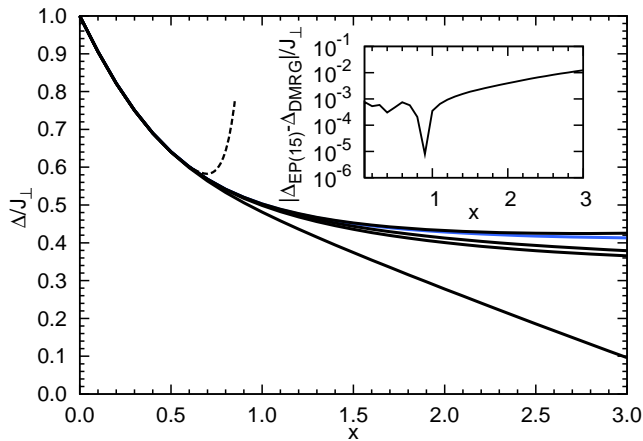}
 \caption{(Color online) Gap $\Delta(x)$ of the uniform spin ladder vs.\ relative leg coupling $x$ for various orders 
 ($6, 10, 14, 15$) using the direct evaluation (black, solid), perturbative evaluation (plain series; order 15; black, dashed) and a $\nicefrac{1}{L^2}$ finite-size scaling DMRG result (dark gray (blue), solid).  The lowest curve in direct evaluation at $x=3$ stems from order 6;
the highest curve from order 15. The 15$^{th}$ order curve in direct evaluation agrees very well
 with the DMRG results. The deviations to the DMRG results are depicted in the inset.}
 \label{plot:Gap_ladder}
\end{figure}

The one-triplon dispersion is calculated up to order 15. The dispersion is obtained by a Fourier transform of the one-triplon sector of $H_{\text{eff}}$. The dispersion reads as
\begin{align}
\omega(k,x)=t_0+\sum_{d=1}^{n}2t_d\cos(dk)\,\text{,}
\end{align}
where $t_d$ is a hopping element over the distance $d$.
Figure \ref{plot:Dispersion_ladder} shows the dispersion for various values of $x$. The plain
series of the perturbative evaluation is depicted as dashed light gray (orange) line. The coefficients of this series (see Table\ \ref{tab:coefficients_dispersion}) agree quantitatively with other series expansion results\cite{oitma96b} up to order 8.

The coefficients of the perturbative series of the spin gap match those of other series expansion 
methods \cite{Zheng1998} up to order 13 . The plain series is reliable up to $x\approx0.6$.
For larger values of $x$ extrapolations are needed. For the dispersion we used the extrapolation scheme based on a re-expansion of the original series in terms of a suitable internal parameter \cite{schmi03a}
 $p(x)$ 
\begin{align}
\label{eq:internal_parameter}
p(x)=1-\frac{\Delta(x)}{(1+x)J_{\bot}}
\end{align}
where $\Delta(x)$ denotes the gap. In order to use the above mapping $x\to p$ 
a reliable extrapolation of the gap $\Delta(x)$ is needed. This was achieved by dlog-Pad\'e
extrapolations which work very robustly for the gap.
 The dashed dark gray (blue) lines in Fig.\ \ref{plot:Dispersion_ladder} represent the plain series in this internal parameter without any further extrapolation.

The solid black lines are the results of the direct evaluation. For small $x$ the 
perturbative and the direct evaluation agree very well. 
The direct results are again very robust for larger $x$ as well.

To corroborate that the deepCUT results are reliable even for relatively large values of $x$
its spin gap is compared to the one obtained in DMRG \cite{Raas2011,ALPS_1,ALPS_2} in 
Fig.\ \ref{plot:Gap_ladder}. The solid black lines represent the deepCUT results. 
The solid dark gray (blue) line depicts the results of a $\nicefrac{1}{L^2}$ finite-size scaling of the DMRG results based on the ansatz
\begin{align}
\label{eq:ansatz_Gap_DMRG}
\Delta\left(z\right)=\Delta(\infty)+a_1z+a_2z^{\nicefrac{3}{2}}+a_3z^2
\end{align}
with $z:=\nicefrac{1}{L^2}$.

The deepCUT data shows that for larger $x$ a higher order is needed to compute the gap 
accurately. This is understood on the basis of the correlation length of the system.
A larger value of $x$ enables us to capture the physics of systems with
larger correlation length $\xi$, which is given by
$\nicefrac{v}{\Delta}$ where $v$ is the spin-wave velocity in absence of a gap.
By construction of the (de)epCUT, the order $n$ defines the range of processes which
are still captured. Hence, one can expect a reliable result as long as 
\begin{equation}
n\gtrapprox \xi
\qquad
\Leftrightarrow
\qquad
 n \gtrapprox\nicefrac{v}{\Delta}(x),
 \label{eq:condition}
\end{equation}
where the lattice constant is set to unity. The velocity $v$ can be estimated by  fitting
$v\sin(k)$ to the maximum of the dispersion. We find indeed that \eqref{eq:condition}
is satisfied up to $x\approx 3$ for $n=15$.
The deepCUT curve for $n=15$ agrees well with the DMRG results. The deviations shown in the inset are rather small. For $x=2$, it is below $10^{-2}J_\perp$. 
Furthermore, the dispersions shown in Fig.\ \ref{plot:Dispersion_ladder} agree with exact diagonalization results. \cite{brehm99}

We conclude that the reliable results for the uniform spin ladders beyond $x=1$ illustrate the efficiency of the deepCUT approach for a model with an equidistant spectrum.

\subsection{Spectral weights}

\begin{figure*}
 \begin{minipage}{0.49\textwidth}\includegraphics[width=\textwidth]{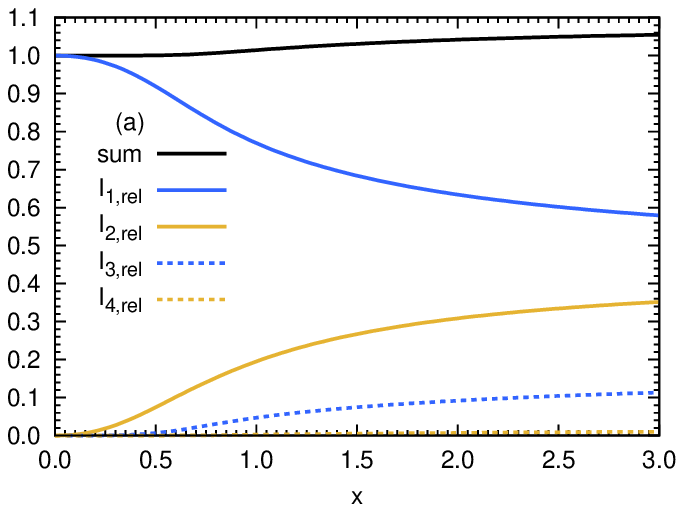}\end{minipage}
 \begin{minipage}{0.49\textwidth}\includegraphics[width=\textwidth]{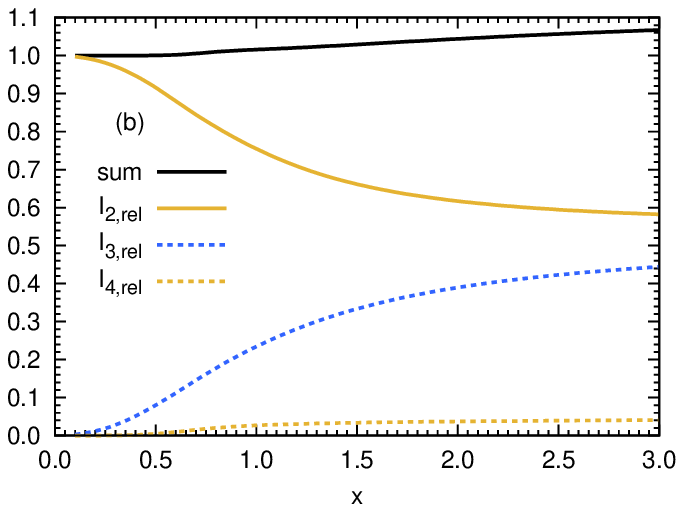}\end{minipage}
 \begin{minipage}{0.49\textwidth}\includegraphics[width=\textwidth]{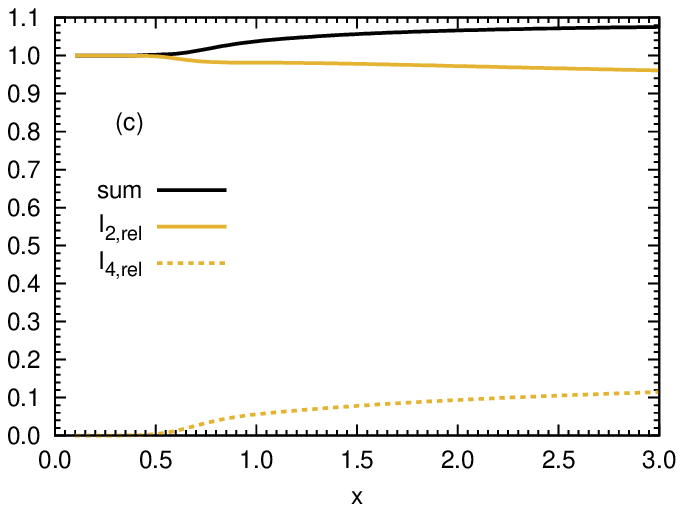}\end{minipage}
 \begin{minipage}{0.49\textwidth}\flushleft
 \caption{\label{plot:spectral_weights}
 \hspace{42cm}}
 (Color online) Spectral weights of the uniform spin ladder for different observables defined in
 	Eq.\ \eqref{eq:observ_def} and numbers of triplons vs.\
  relative leg coupling  $x$. Panel (a) shows the $S=1$ observable $O^{\text{I}}$; 
  panels (b) and (c) show the parallel and perpendicular $S=0$ observables $O^{\text{II}}$ 
  and $O^{\text{III}}$. 
  The calculations were carried out to order $8$ for the modified observables
  $x\cdot O^{\text{I}}$, $x\cdot O^{\text{II}}$, and $x\cdot O^{\text{III}}$ 
  using deepCUT with the generator scheme ${\widehat \eta_{2}}$.
 \end{minipage}
\end{figure*}

Here, we use the transformation of self-adjoint observables $O$ (cf.\ Sec.\ \ref{struct:observables}) to address the issue of spectral weights.
We denote the subspace spanned by the states with $q$ quasiparticles  by $\text{QP}_q$.
As in previous work, \cite{knett01b,schmi03c,schmi04a,schmi05b} 
we split the total spectral weight at zero temperature
into its contributions from the different subspaces $\text{QP}_q$:
\begin{subequations}
\label{eq:spec_weight_def}
\begin{align}
 I_q & := \sum\limits_{\left|i\right\rangle \in \text{QP}_q}\left|\left\langle i\right|O\left|0\right\rangle\right|^2
 \\
 &= \left\langle0\right|O^0_qO^q_0\left|0\right\rangle,
\end{align}
\end{subequations}
where $O^q_p$ stands for the sum over all terms of the transformed observable
consisting of $q$ creation operators  and $p$ annihilation operators in normal ordering.
The state $\left|0\right\rangle$ denotes the vacuum state of the effective model, i.e., the ground-state of the Hamiltonian.

If the subspaces $\text{QP}_q$ have been separated by the CUT, i.e., the effective
Hamiltonian does no longer mix them, the spectral weights defined by \eqref{eq:spec_weight_def}
coincide with the ones defined previously. \cite{knett01b,schmi05b}
The spectral weights correspond to the integral over momentum and frequency of
the corresponding dynamic structure factor $S_q(\textbf{k},\omega)$ where the
subscript $q$ denotes the contribution of the subspace $\text{QP}_q$.
Thus, \emph{separate} sum rules exist for each $\text{QP}_q$.
Such a split-up is only possible 
because the dynamics does not mix the subspaces according
to the above assumption.
We recall that the dynamic structure factors encode the response of
various inelastic scattering experiments. 

If the subspaces $\text{QP}_q$ are not or not all separated, the equal-time definition
\eqref{eq:spec_weight_def} is still well defined.
But, $I_q$ can no longer be interpreted as the sum rule of $S_q(\textbf{k},\omega)$
because the subspaces mix in the course of the dynamics induced by the Hamiltonian. 
Nevertheless, the values $I_q$ provide a plausible measure
of the importance of the subspaces of different number of excitations.
A large spectral weight for low numbers of quasiparticles indicates that results of scattering experiments can be understood from the spectral densities involving only low numbers of quasiparticles.

In the this work, we concentrate on the spectral weights for the observables
\begin{subequations}
\label{eq:observ_def}
\begin{align}
 O^{\text{I}}  &=S^{\text{L},z}_0,\\
 O^{\text{II}} &=\textbf{S}^{\text{L}}_0\cdot\textbf{S}^{\text{L}}_1,\\
 O^{\text{III}}&=\textbf{S}^{\text{L}}_0\cdot\textbf{S}^{\text{R}}_0
\end{align}
\end{subequations}
to illustrate the transformation of observables. The observable $O^{\text{I}}$ induces
a local spin $S=1$ excitation which can be studied experimentally by inelastic neutron
scattering. \cite{notbo07,hong10b,schmi12}
The observables $O^{\text{II}}$ and $O^{\text{III}}$ induce $S=0$ excitations
which can be studied by optical probes, e.g., Raman scattering 
\cite{schmi01} or infrared absorption, \cite{windt01,nunne02} in polarizations parallel and 
perpendicular to the ladder, respectively. 
Because triplons are $S=1$ states, both observables $O^{\text{II}}$ and $O^{\text{III}}$
induce no contributions in the one-triplon channel: $I_1=0$.
The calculation of the corresponding spectral densities $S_q(\textbf{k},\omega)$ is left to
future studies. 

Since the description in terms of triplons on rungs is obviously best 
for low values of $x$, we expect that more and more triplons need to be addressed
upon increasing $x$.
To assess the relative importance of different triplon channels, we introduce 
the relative weights $I_{q,\text{rel}}=\nicefrac{I_q}{I_\text{tot}}$. They can be calculated 
using the sum rule
\begin{align}
 I_{\text{tot}}:=&\sum\limits_{q=1}^{\infty} I_q =
 \left\langle0\right|OO\left|0\right\rangle-\left\langle0\right|O\left|0\right\rangle^2
 \label{eq:sumrule}
\end{align}
for the total spectral weight $I_\text{tot}$. 
For the observables defined in \eqref{eq:observ_def}, the total weights are given by
\begin{subequations}
 \begin{align}
  I_\text{tot}^{\text{I}}  &= \frac{1}{4},\\
  I_\text{tot}^{\text{II}} &= \frac{3}{16} - \frac{Y}{4}-\frac{Y^2}{4},\\
  I_\text{tot}^{\text{III}}&= \frac{3}{16} - \frac{Z}{2}-Z^2
 \end{align}
\end{subequations}
with the variables
\begin{subequations}
\begin{align}
Y&:=          2 \left\langle0\right|\textbf{S}^{L}_0\cdot\textbf{S}^{L}_1\left|0\right\rangle=
\frac{\partial E_0}{\partial x},
\\
Z&:= \phantom{2}\left\langle0\right|\textbf{S}^{L}_0\cdot\textbf{S}^{R}_0\left|0\right\rangle=
E_0-x\cdot\frac{\partial E_0}{\partial x},
\end{align}
\end{subequations}
where we use the ground-state energy per rung $E_0$.

We focus on the spectral weights in the first four-triplon (four-quasiparticle) 
channels $I_1, I_2, I_3$, and $I_4$ up to large values of the relative leg coupling $x=3$ using deepCUT. 
In this region, a complete decoupling of all subspaces using ${\widehat \eta_\text{pc}}$ or ${\widehat \eta_{4}}$ is no longer possible because divergences occur in the numerical integration of the flow.
This problem is well-known from sCUT; it stems from the overlap of continua of different quasiparticle number \cite{Fischer2010,Duffe10,Duffe11}. In this situation, the sorting of quasiparticle spaces ascending by energy is no longer possible.
In the perturbative evaluation of epCUT, no divergencies appear if
the energies in $H_0$ are separated and indeed ordered according to
ascending number of quasiparticles 
because the hierarchy in Eq.\ \eqref{eq:flow_perturbative} precludes any feedback of high-order coefficients to low-order coefficients. In the alternating spin ladder, the epCUT based on the
quasiparticle number must be modified for $y \ge 3$. But, we stress that this reflects
a more sophisticated physics which must be considered in the choice of the generator.
It does not represent a conceptual problem of epCUT.

To avoid convergence problems due to overlapping continua in deepCUT,
we aim only at decoupling subspaces with at most 
two quasiparticles using ${\widehat \eta_{2}}$ while keeping monomials linking subspaces with higher number of quasiparticles, for instance, $\text{QP}_3 \leftrightarrow \text{QP}_5$.
As a consequence, the observables are transformed to a quasiparticle basis where three and four quasiparticle states still couple to other subspaces. 

The spectral weights for the $S=1$ observable $O^{\text{I}}$ depicted  of Fig.\
\ref{plot:spectral_weights}(a) agree well with pCUT results \cite{knett03b,schmi05b,schmi04b} for
small values of $x$. Note that only the one- and the two-quasiparticle channel can be compared
quantitatively because the pCUT separated also the three- and four-quasiparticle subspaces,
but the present calculation does not.

For $O^{\text{I}}$, most weight is concentrated in the first two quasiparticle channels. 
Even at $x=3$, the one-triplon channel still contains $57.9\%$ of the total weight. The relative weight of the two- and three-triplon channel rises up to $35.2\%$ and $11.3\%$, respectively. 
The four-triplon weight remains negligible. The sum rule is slightly violated because
the accumulated relative weights exceed $100\%$. This inaccuracy is related to the finite order
of calculation. The degree of the violation of the sum rule can be used as a measure for the
 reliability of the results. Even at $x=3$, the excess weight is only $5.5\%$.

Figure\ \ref{plot:spectral_weights}(b) shows the spectral weights $I_q$ for the $S=0$ observable $O^{\text{II}}$. Since triplons have spin $S=1$, there cannot be any weight in 
the one-triplon channel. Instead, most weight is concentrated in the two-triplon channel which
agrees well with pCUT results \cite{knett01b,knett03b,schmi04b,schmi05b}. Compared to
$O^{\text{I}}$, the three-triplon channel is much more pronounced displaying a relative weight
of $44.4\%$ at $x=3$. At this value, the sum rule is fullfilled within $6.7\%$.

The observable $O^{\text{III}}$ is symmetric with respect to the ladder's centerline.
This implies that this observable does not change the parity of a state 
\cite{knett01b,schmi05b}. A single triplon is an odd excitation with respect to the
ground-state. Hence $O^{\text{III}}$ can create or annihilate triplons only in pairs.
There is no weight in odd channels in Fig.\ \ref{plot:spectral_weights}(c). 
As a consequence, the spectral weight is distributed over the two- and four-triplon channels only. 
Our results do not indicate a sizable contribution from six and more triplons,
but this has not been studied quantitatively.
At $x=3$, the sum rule is violated by $7.5\%$. Indeed, this violation sets in
at about $x=0.6$ when the four-triplon weight becomes significant. Thus we presume that the
latter is a bit overestimated, but we could not identify the mechanism for this effect.
The perturbative results for the weights fulfill the sum rule to the required order.

\section{Results for alternating spin ladder}

\subsection{Ground-state energy}

\begin{figure}
 \includegraphics[width=\columnwidth]{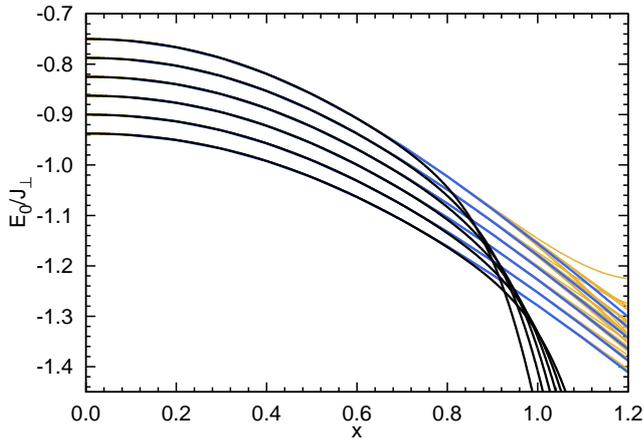}
 \caption{(Color online) Ground-state energy per rung 
  $E_0$ of the alternating spin ladder vs.\ relative leg coupling $x$ for various
  values of $y\in\{1,1.1,1.2,1.3,1.4,1.5\}$ and various evaluation techniques based on the DES in
  order 16. The highest curve at $x=0$ is $y=1$ and the lowest curve is $y=1.5$. Again the direct
  evaluation [dark gray (blue) line] yields a much more stable and reliable extrapolation of the
  plain series (black line) than the various Pad\'{e} extrapolants (orange line).}
 \label{plot:E_0_alt_ladder}
\end{figure}

\begin{figure}
 \includegraphics[width=\columnwidth]{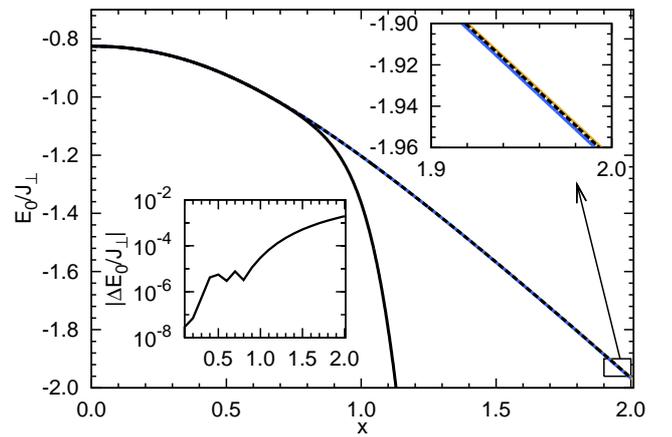}
 \caption{(Color online) Ground-state energy per rung $E_0$ of the alternating spin ladder vs.\ relative leg coupling $x$ for $y=1.2$ for various methods. Depicted are the results of the perturbative evaluation (order 16; black, solid line) the direct evaluation (order 16; black, dashed line), a high-level sCUT calculation [$d$=(12,10,10,6,6,5,5,4,4), light gray (orange), solid line] and a DMRG calculation (dark gray (blue), solid line). The direct evaluation agrees very well with the sCUT and the DMRG results. The deviation $|\Delta E_0|=|E_{0,\text{direct}}-E_{0,\text{DMRG}}|$ between the results of the DMRG and of the direct evaluation is shown in the lower inset.}
 \label{plot:E_0_alt_ladder_comparision}
\end{figure}

For $J_{\bot}^{\text{o}}\neq J_{\bot}^{\text{e}}$, the ground-state energy is calculated up to order 16. Due to the doubled unit cell, only a slightly lower order can be reached
than for the uniform spin ladder. Roughly, we need double the number of coefficients for the alternating
 spin ladder. The ground-state energy per rung is given by $E_0=\nicefrac{h_0}{2}$.

The perturbative results from  epCUT are shown in Fig.\ \ref{plot:E_0_alt_ladder}. As expected
the ground-state energy decreases upon rising $y$. The black lines represent the results of the 
plain series for various $y$. The coefficients are given in Table\
 \ref{tab:coefficients_gse}. The light gray (orange) lines correspond to various Pad\'{e}-extrapolants. 
The plain series is reliable up to $x\approx 0.75$ for $y=1$ and up to $x\approx 0.85$ for $y=1.5$. 
So, the $x$ up to which the series is reliable depend on the value of $y$.
Since a larger value of $y$ supports the dominance of the unperturbed Hamiltonian
$H_0$ it is clear that an increasing $y$ supports the validity of the perturbation.
The dark gray (blue) lines represent the results of the deepCUT. These results again represent 
a very robust extrapolation of the perturbative results up to larger $x$.

To show the efficiency of the epCUT, the results for the ground-state energy per rung are compared
to the results of an sCUT calculation and a DMRG (Refs.\ \onlinecite{Raas2011,ALPS_1,ALPS_2}) calculation. 
The sCUT was performed with the ground-state generator and a $d$=(12,10,10,6,6,5,5,4,4) truncation. The DMRG results ($L=20,40,\dots,100$, $m=100$) are extrapolated according to Eq.\ \eqref{eq:ansatz_DMRG}.
Figure \ref{plot:E_0_alt_ladder_comparision} compares the results of the various approaches. 
They agree very well with one other. The deviations between the results of the direct evaluation and the DMRG calculation are small (see upper inset). 
For $x=1.5$, the deviation is less than $10^{-3}J_\perp$.

\subsection{Dispersion}

\begin{figure*}
 \includegraphics[width=\columnwidth]{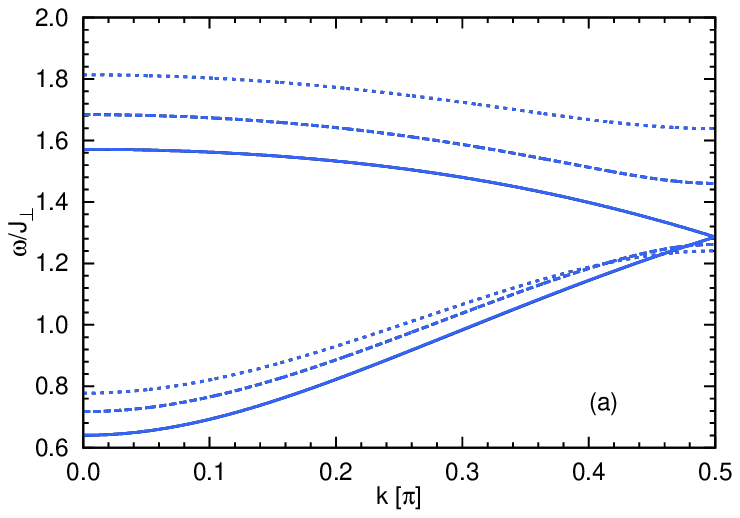}
 \includegraphics[width=\columnwidth]{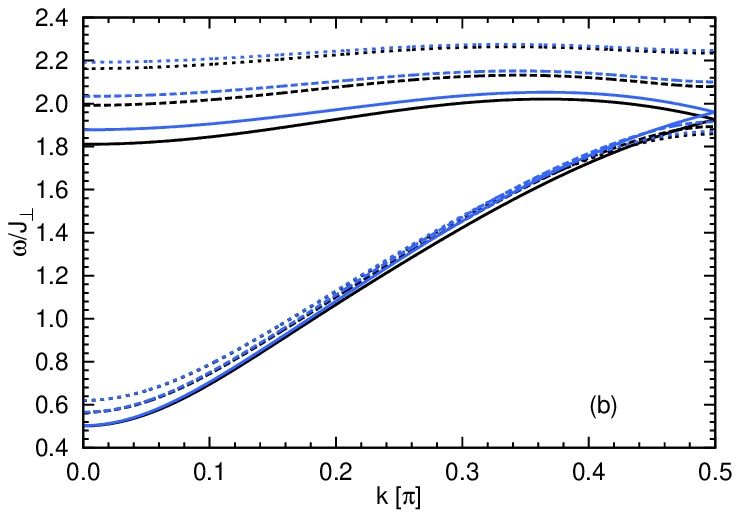}
  \caption{(Color online) 
  (a) Dispersion $\omega(k)$ of the alternating spin ladder 
  for $x=0.5$ and various values of
   $y\in\{1,1.2,1.4\}$ (solid, dashed, dotted lines; order 13). 
   The direct evaluation is depicted by dark gray (blue) lines 
   and the perturbative one by black lines which are hardly visible because
   they are just below the other lines. 
   For the perturbative results the plain series are used. 
   (b) Dispersion $\omega(k)$ of the alternating spin ladder for $x=1$ and various values of $y\in\{1,1.2,1.4\}$ 
   (solid, dashed, dotted lines; order 13). The direct evaluation is depicted by dark gray (blue) lines 
   and the perturbative one by black lines. 
   For the perturbative results, the plain series in the internal parameter $p_\text{a}(x)$
   defined in Eq.\ \eqref{eq:pa_def}  without any further extrapolation is used. 
  The upper branches differ slightly, but in general both evaluation techniques agree well.}
  \label{plot:alt_Dispersion}
\end{figure*}

\begin{figure}
 \includegraphics[width=\columnwidth]{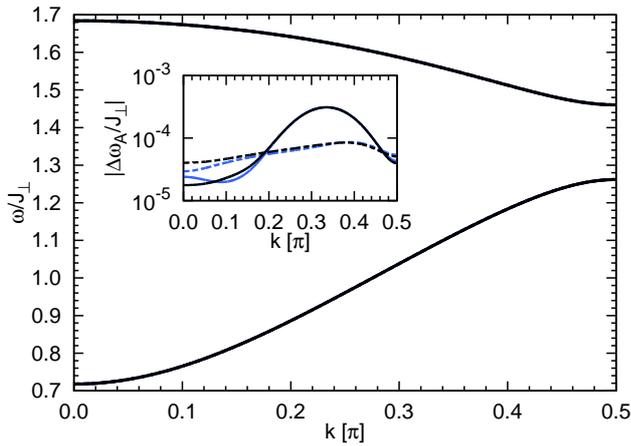}
 \caption{(Color online) Dispersion $\omega(k)$ of the alternating spin ladder for $x=0.5$ and $y=1.2$. Displayed are
 the direct evaluation (order 13; black, dashed line) and the perturbative one (order 13;
    black, solid line), and an sCUT result ($d$=(12,10,10,6,6,5,5,4,4), dark gray (blue), solid). 
    All dispersions lie on the top of each other. The deviations
    $|\Delta\nicefrac{\omega}{J_{\bot}}| = |\nicefrac{\omega_{\text{direct}}}{J_{\bot}} -
    \nicefrac{\omega_{\text{pert}}}{J_{\bot}}|$ (black) and $|\Delta\nicefrac{\omega}{J_{\bot}}|
    = |\nicefrac{\omega_{\text{sCUT}}}{J_{\bot}} - \nicefrac{\omega_{\text{pert}}}{J_{\bot}}|$
    (blue(dark gray)) are shown in the inset. The solid (dashed) lines depict the lower (higher)
     branch.}
\label{plot:Dispersion_alt_ladder_comp}
\end{figure}

\begin{figure}
 \includegraphics[width=\columnwidth]{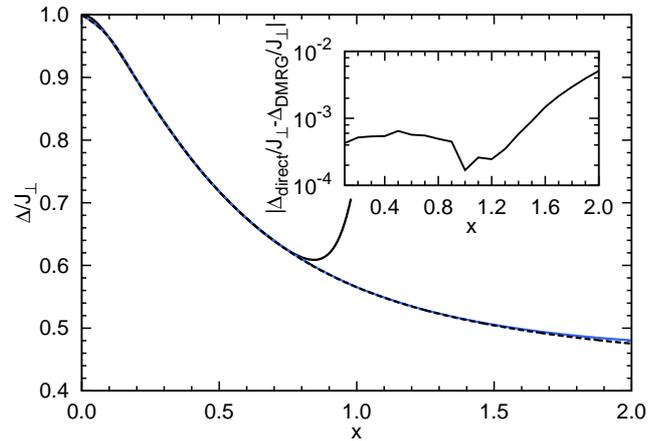}
 \caption{(Color online) Gap $\Delta(x)$ of the alternating spin ladder vs.\ relative leg coupling $x$ for $y=1.2$ from the
  pertubative evaluation (order 13; black, solid line), the direct evaluation (order 13; black, dashed line)
   and a DMRG result (dark gray (blue), solid line) extrapolated to the thermodynamic limit $L=\infty$
   by finite-size scaling $\propto \nicefrac{1}{L^2}$. The results agree well. The deviations
   between deepCUT and DMRG results are plotted in the inset.}
 \label{plot:Gap_alt_ladder_comp}
\end{figure}

The dispersion is calculated up to order 13 for the alternating ladder. An important step
is the Fourier transform of the hopping in the one-triplon sector of $H_{\text{eff}}$. 
But, the doubled unit cell has to be taken into account which halves the Brillouin zone. In
 return, the dispersion acquires two branches reading as
\begin{align}
\omega_{\pm}(k)=\frac{M_{\text{ee}}+M_{\text{oo}}}{2}
\pm \sqrt{\frac{(M_{\text{ee}}-M_{\text{oo}})^2}{4}+M_{\text{eo}}^2},
\end{align}
where $M_{ij}$ stands for the Fourier transform of the hopping processes 
from a rung of parity $i$ to a rung of parity $j$.

Figure \ref{plot:alt_Dispersion}(a) displays the dispersions for $x=0.5$ and various values of $y$ (see also Table\ \ref{tab:coefficients_dispersion}). The solid lines represent $y=1$, the dashed ones $y=1.2$ and the dotted ones $y=1.4$. The dark gray (blue) lines stand for directly evaluated results and the black lines for the
 perturbatively evaluated ones. For $x=0.5$ the plain series is used. Both results agree very well. For $y=1$ we retrieve the uniform ladder and the two branches meet at
$k=\nicefrac{\pi}{2}$. As expected the branches split at $k=\nicefrac{\pi}{2}$ 
once $y>1$ holds due to the reduced translational symmetry. 

To show the efficiency of the epCUT, the dispersion relations for $y=1.2$ are compared to the dispersion from an sCUT calculation in Fig.\ \ref{plot:Dispersion_alt_ladder_comp}. The sCUT was performed with the generator $\eta_1$ and the truncation $d$=(12,10,10,6,6,5,5,4,4). The
dispersions match perfectly. The deviation between sCUT and the perturbative evaluation is less than $10^{-3}J_\perp$. The differences in the upper branch are larger than those in the 
lower branch.

Furthermore, the gap for $y=1.2$ is compared to the gap obtained by a DMRG calculation. \cite{Raas2011,ALPS_1,ALPS_2} The finite-size scaling is carried out again based on Eq.\ \eqref{eq:ansatz_Gap_DMRG}. In Fig.\ \ref{plot:Gap_alt_ladder_comp}, the solid
black line shows the perturbative result and the dashed black line the result of  the
deepCUT. The result of the DMRG calculation is depicted as solid dark gray (blue) line. The deviations between the deepCUT and the DMRG calculation are shown in the inset.
 Again the results agree very well, e.g., the deviation is less than $10^{-2}J_\perp$ even at
 $x=2$ .

The dispersions at $x=1$ of the direct and of the perturbative evaluation are plotted in Fig.\ \ref{plot:alt_Dispersion}(b). The perturbative results are rendered using the plain series in an
internal parameter. \cite{schmi03a} The  parameter, however, defined in Eq.\
 \eqref{eq:internal_parameter} does not work because at $y\neq 1$ it behaves like
 $p(x) \propto x^2$ and not linearly in $x$. Thus, the series in $x$ can not
 be re-expressed in a series in $p$.  So, we modify the internal parameter
\begin{align}
\label{eq:pa_def}
p_\text{a}(x)&:=1-\frac{1}{1+y}
\cdot\left(M_{\text{ee}}+M_{\text{oo}}-2|M_{\text{eo}}|\right),
\end{align}
where all matrix elements are taken at vanishing wave vector $k=0$.
We choose this parameter because it reproduces the previous definition
 \eqref{eq:internal_parameter} for $y=1$ and for $x\to\infty$.
In addition, it fulfills  $p_\text{a}\propto x$ for $x\to 0$ for all values of $y$. 
Otherwise, the extrapolation can be performed as before. \cite{schmi03a}
The Fourier-transformed matrix elements $M_{ij}(x)$ are obtained by robust dlog-Pad\'e
extrapolations.
The results of deepCUT and of the series in this internal parameter agree very well.

The epCUT results for the alternating ladder exemplify the efficiency of this CUT for 
a system with a \emph{non-equidistant} spectrum in $H_0$. 
Thereby, the range of applicability of perturbation by CUTs is crucially enhanced because
the previous pCUT (Refs.\ \onlinecite{uhrig98c,knett00a}) is restricted to equidistant unperturbed spectra.

\section{Conclusions}

In this article, we presented a methodological development and illustrated it
for a well-understood model. We extended the previously known
perturbative continuous unitary transformation (pCUT) in two ways. 

First, we formulated the perturbative realization of the CUTs directly in 
second quantization. Thereby, the unperturbed part is no longer
restricted to an equidistant spectrum of energy eigen values.
The direct expansion of all coefficients in the effective Hamiltonian
 is not efficient enough. But, by tracking the
powers in the expansion parameter $x$ of all the physical processes, it is
possible to identify the relevant ones for the low-energy effective model:
ground-state energy, single-quasiparticle dispersion, and two-quasiparticle
interactions. We could show that this leads to an efficient and competitive
approach to obtain effective models. Their parameters are computed as series
in the expansion parameter. For distinction, we baptized the enhanced approach
enhanced perturbative CUT (epCUT).

Second, we found that the system of differential flow equations, which has been reduced
to provide the perturbative series representation of the effective model, can also 
be directly evaluated. It appears that this directly evaluated perturbative CUT (deepCUT)
yields a very robust and reliable way to exploit the information
in the perturbative differential flow equations. In some sense, one can think of
it as a robust extrapolation although we stress that it is not an algorithm
applied to a series. The deepCUT provides the parameters of the effective
models for given initial Hamiltonian. Each set of initial parameters requires
a numerical integration of the flow equations which is a moderate numerical task.
The essential effort lies in deriving the system of differential flow equations
which is the same as for the epCUT.

The epCUT and the deepCUT are illustrated by the very simple
model of a perturbed harmonic oscillator where all equations can be 
written explicitly. Thereby, an example with infinite-dimensional 
local Hilbert space is given, although the unperturbed spectrum
is equidistant for the sake of simplicity. The equidistance is not an essential point since two coupled harmonic oscillators with differing eigenenergies coupled by quartic terms would constitute another straightforward example of only slightly higher complexity, but with non-equidistant unperturbed spectrum.
The equidistance is not an essential point since two coupled harmonic
oscillators with differing eigenenergies coupled by quartic terms
would constitute another straightforward example of only slightly
higher complexity, but with non-equidistant unperturbed spectrum.

Both abstract key results were also illustrated by calculations for 
antiferromagnetic $S=1/2$ spin ladders with two legs. Two types of
spin ladders were studied. The expansion parameter is the leg coupling
relative to the (smallest) rung coupling. The uniform spin ladder with the same value
of the rung coupling is the standard model which is very well studied.
The alternating spin ladder with alternating rung couplings has not yet been 
studied to our knowledge. For the present purposes, it constitutes a model 
with a non-equidistant unperturbed spectrum if the perturbation is set
up around the rung Hamiltonian.

For the uniform spin ladder, the known series coefficients could be 
retrieved by epCUT. The corresponding results for the alternating spin ladder
have not been published elsewhere. They show that general unperturbed
spectra can be treated by epCUT.

The data obtained by deepCUT illustrate that this approach yields surprisingly
robust results. The uniform spin ladder could be treated up into the strong-leg limit
with values of the relative leg coupling $x=J_\parallel/J_\perp$ of up to 
$x=3$.
This is a parameter regime which was not accessible by CUTs before. \cite{schmi05b,notbo07}

The limit of the applicability of deepCUT
 can be understood in terms of the correlation length.
A deepCUT calculation in order $n$ in a perturbation linking adjacent
sites allows us to capture processes up to the range $n\cdot a$ where
$a$ is the lattice constant. Hence, reliable results can be expected if
the correlation length $\xi=v/\Delta$ is lower than $n\cdot a$.
The deepCUT results for the alternating ladder are also very robust,
although a little less than for the uniform ladder.

Further work on the precise preconditions required for the applicability
of epCUT and deepCUT is called for. Also, their applicability 
 to two- or higher-dimensional systems deserves to be 
studied in the future.

 The deepCUT approach works on the level of monomials of creation and
 annihilation operators, i.e., in second quantization. Thus, essentially all
 symmetries of the lattice problem under study can be preserved by construction.
 A large variety of the generators can be realized. 

In a nutshell, we advocate two approaches
to derive effective models in a systematically controlled way in this article.
They have been illustrated for a perturbed harmonic oscillator
and spin ladders, and we expect that applications to
many other models will soon be possible.

\begin{acknowledgments}
We gratefully acknowledge many useful discussions with Sebastian Duffe, Tim Fischer,
Mohsen Hafez, Carsten Raas, and Kai P. Schmidt. Our research relies on funding
by the NRW-Forschungsschule ``Forschung mit Synchrotronstrahlung in den Nano- und 
Biowissenschaften'' (N.A.D.) and by the Mercator Stiftung (H.K.).
\end{acknowledgments}

\appendix

\section{Extended A Posteriori Simplification Rule}
\label{struct:extended_a-posteriori}

\begin{figure*}
 \includegraphics[width=\textwidth]{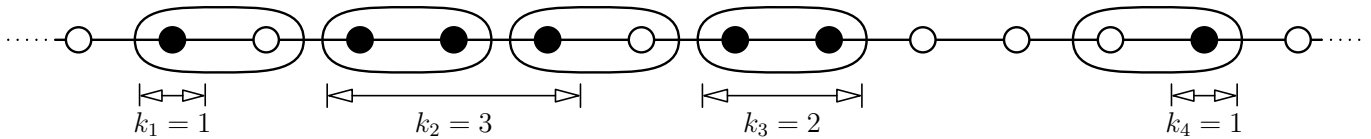}
 \caption{Decomposition of the sites with creation operators 
 (or the annihilation operators, respectively) of a monomial into
linked subclusters $k_i$ and its covering with first-order generator terms. 
Each circle stands for a rung of the spin ladder.
Filled circles represent rungs where the local action of the monomial differs from identity. 
At most, two adjacent local operators can be canceled by a single commutation with $\eta^{(1)}$; 
this is represented by ellipses.}
 \label{img:extended_rule}
\end{figure*}

The upper bound ${{\widetilde{O}}_{\max}}$ for the maximum order can be reduced by considering the real-space structure
of the monomial. For clarity, we restrict ourselves to one-dimensional models. As for the basic
simplification rule, we discuss the effect of commutations with first-order terms present in the initial Hamiltonian. This is sufficient because any more complicated monomials in the generator
have been induced by commutations of a number of first-order terms. Hence, their gain in number
of involved quasiparticles is paid for by a correspondingly higher order in $x$. Thus, one may
safely restrict the consideration to the basic building blocks present in the initial Hamiltonian.

For the spin ladder in terms of triplon operators [Eq.\ \eqref{eq:spinhammi}],
a commutation with the generator $\eta^{(1)}$ cancels at most two local creation or annihilation operators on \emph{adjacent} sites. Therefore, sparse and extended monomials require more commutations in
order to reduce their local operators compared to monomials with the same numbers of operators 
which are more localized in real space.

At first, we study the ground-state energy per rung, i.e., the coefficient of the identity operators
summed over all rungs, in highest order. The clusters of the creation and of the annihilation operators
are treated separately. Both are decomposed into linked subclusters of size $k_i^c$ and $k_i^a$ (see
Fig.\ \ref{img:extended_rule}). To cancel all local operators, each subcluster needs to be covered by
$\left\lceil {\frac{k_i}{2}} \right\rceil$ first-order generator terms. In conclusion,
 \begin{align}
  K_0=\sum\limits_i \left\lceil { \frac{k_i}{2}} \right\rceil\label{eq:plain_cluster_sum}
 \end{align}
commutations with $\eta^{(1)}$ are needed for the clusters of creation or annihilation operators
to be reduced to the coefficient of the identity operator. 
This argument leads to the extended upper bound for the maximum order
\begin{align}
\label{eq:omaxbound_app1}
 {{\widetilde{O}}_{\max}} = n-K_0^c-K_0^a.
\end{align}
For a single linked cluster, this formula resembles the result obtained for the basic simplification rule \eqref{eq:basic_a-posteriori}.

The formula \eqref{eq:omaxbound_app1} can be generalized to 
\begin{align}
\label{eq:omaxbound_general}
 {{\widetilde{O}}_{\max}}=n-K_q^c-K_q^a
\end{align}
for the subspace $\text{QP}_q$ of states with $q$ quasiparticles leading to modified cluster sums $K_q$.
Let $q$ be the number of the targeted subspace with the highest number of quasiparticles.
This means that $q$ ist the maximum number of local creation and annihilation operators allowed in a monomial targeted up to order $n$. 
Terms which affect more quasiparticles have to be reduced to affecting at most $q$
quasiparticles by commutations with $\eta^{(1)}$ until at most $q$ local creation and annihilation operators are left. To obtain an upper bound ${{\widetilde{O}}_{\max}}$, one has to choose $q$ positions for local operators to be kept in the cluster in such a way that the other creation and/or annihilation 
operators can be canceled by a minimum number of commutations.  To this end, one also
has to consider that the commutations with hopping terms stemming from $H^{(1)}$ may also
shift creation and/or annihilation operators so that they form adjacent pairs which
can be canceled by pair creation or annihilation. But, it turns out that this mechanism can reduce the cluster sum $K_0$ \emph{at most} by unity, while the elimination of a pair of adjacent local operators \emph{always} reduces the cluster sum by unity. 
Hence, the latter process dominates and provides the correct upper bound ${{\widetilde{O}}_{\max}}$.

For the hopping in the symmetric ladder model, the above approach
means to select sites at the edges of odd subclusters first. 
This saves one commutation for each local operator kept. Let
$\alpha$ be the number of odd clusters. The cluster sum $K_0$ is reduced in this way by
\begin{align}
 d_1 = \min(\alpha,q).
\end{align}
If more local operators remain, i.e., $\alpha < q$, the most 
efficient way to place them is in pairs on even subclusters.
This reduces the cluster sum additionally by
\begin{align}
 d_2 = \left\lfloor { \frac{q-d_1}{2}} \right\rfloor,
\end{align}
where $\left\lfloor {y} \right\rfloor$ is the largest integer which is still smaller or
equal to $y$.

In conclusion, the cluster sums are reduced when one is aiming at higher quasiparticle subspaces
according to
\begin{subequations}
\begin{align}
K_q^{\prime}=K_0 - d_1 - \left\lfloor {\frac{q-d_1}{2}} \right\rfloor=K_0-\left\lfloor {\frac{q+d_1}{2}} \right\rfloor.
\end{align}
To avoid unreasonable negative results, this expression has to be checked against zero 
to obtain the final result
\begin{align}
 K_q=\max(K_q^{\prime},0).
\end{align}
\end{subequations}

We remark that the extended simplification rule can be easily adapted to other models with 
monomials of first order in the generator to create or annihilate an arbitrary number ${\Delta \text{QP}}$ of quasiparticles on adjacent sites. 
For further refinements of ${{\widetilde{O}}_{\max}}$, one may consider the triplon polarizations $x, y, z$ as well.
But, the derivation and application of an appropriate polarization-sensitive simplification rule 
is beyond the scope of this article, which aims primarily at the proof-of-principle demonstration
of epCUT and deepCUT.

\section{Basic A Priori Simplification Rule}
\label{struct:basic_a-priori}

As stated in Sec.\ \ref{struct:basic_a-posteriori}, the performance of the epCUT algorithm can be
enhanced significantly by avoiding the computation of unnecessary commutators. 
For this purpose, we consider the two normal-ordered products $TD$ and $DT$ in
\begin{align}
 \left[ T,D \right] =TD-DT
\end{align}
separately. Here we discuss $TD$ explicitly; $DT$ is treated in the same way. 
For an analog of the basic simplification rule (Sec.\ \ref{struct:basic_a-posteriori}), 
we estimate the minimum numbers of creation and annihilation operators $c_{TD}$ and $a_{TD}$ which
can appear in the monomials of the normal-ordering of $TD$. We use the numbers $c_T, c_D, a_T$, and $a_D$
from each factor as input. At most
\begin{align}
 s_{TD}=\min(a_T,c_D)
\end{align}
pairs of local operators can cancel in the process of normal-ordering. Hence it follows
\begin{subequations}
\begin{align}
 c_{TD}&\ge c_T+c_D-s_{TD}\\
 a_{TD}&\ge a_T+a_D-s_{TD}.
\end{align}
\end{subequations}

Using these estimates in Eq.\ \eqref{eq:basic_a-posteriori}, one obtains an upper bound 
\begin{align}
\begin{split}
 {{\widetilde{O}}_{\max}} {}_{,TD} = n 
  &- \left\lceil \max\left( \frac{c_T+c_D-s_{TD}}{2}-q,0\right)\right\rceil \\  
  &- \left\lceil \max\left( \frac{a_T+a_D-s_{TD}}{2}-q,0\right)\right\rceil
  \label{eq:basic_a-priori_omax-bound}
\end{split}
\end{align}
with $q$ being the number of the targeted quasiparticle subspace.
Considering also the inverse product $DT$, the commutator $\left[ T,D \right] $ does not need to be calculated while evaluating $\partial_\ell H^{(m)}$ if 
\begin{align}
 m > \max\left({{\widetilde{O}}_{\max}}{}_{,TD},{{\widetilde{O}}_{\max}}{}_{,DT}\right) 
 \label{eq:basic_a-priori_omax-condition}
\end{align}
holds.

As an example, we consider the second order calculation ($n=2$) given in Tabs.\ \ref{tab:list_2nd_order} and \ref{tab:diffeq_2nd_order} for the ground-state energy ($q=0$). Calculating $\partial_\ell H^{(2)}$, the commutator of the monomials
\begin{subequations}
\begin{align}
 T&=t^\dagger_{x,0}t^\dagger_{x,1}\\
 D&=t^\dagger_{y,0}t^\dagger_{y,1}t^{\phantom{\dagger}}_{z,0}t^{\phantom{\dagger}}_{z,1}
\end{align}\end{subequations}
with ${O_\text{min}^{}}(T)={O_\text{min}^{}}(D)=1$ occurs. 
The numbers of local creation and annihilation operators are given by
\begin{subequations}\begin{align}
 c_T&=2 & a_T&=0\\
 c_D&=2 & a_D&=2.
\end{align}
\end{subequations}
In the normal-ordering of  $TD$, no local operator can cancel ($s_{TD}=0$) implying 
$c_{TD}=4$ and $a_{TD}=2$. For the product $DT$, $s_{DT}=2$ pairs of local operators may cancel implying
$c_{TD}\ge 2$ and $a_{TD}\ge 0$. Using Eq.\ \eqref{eq:basic_a-priori_omax-bound}, we find
\begin{subequations}
\begin{align}
 {{\widetilde{O}}_{\max}} {}_{,TD}&=n-\left\lceil 2\right\rceil-\left\lceil 1\right\rceil=-1\\
 {{\widetilde{O}}_{\max}} {}_{,DT}&=n-\left\lceil 1\right\rceil-\left\lceil 0\right\rceil=+1.
\end{align}
\end{subequations}
Since the commutator $[ T,D ] $ yields monomials with a maximum order of at most 1 in the
calculation of $\partial_\ell H^{(2)}$, it can not yield relevant contributions.
Hence it does not need to be evaluated at all.

But in a calculation of order $n>2$ or aiming at a higher quasiparticle subspace $q>0$, Eq.\ \eqref{eq:basic_a-priori_omax-bound} yields higher upper bounds for the maximum order and thus 
the commutator must be evaluated explicitly. 
Note that this basic \emph{a priori} rule is only sensitive to changes of the quasiparticle \emph{numbers}.
It can not anticipate that the commutator in this example actually vanishes due to 
other properties of the hard-core algebra of the triplons.

\section{Extended A Priori Simplification Rule}
\label{struct:extended_a-priori}

The real-space structure of the commutator arguments $T$ and $D$ allows us to extend the above 
a-priori rule in analogy to the extended a-posteriori rule in App.\ \ref{struct:extended_a-posteriori}.
Let $C_T$ and $C_D$ be the clusters of the creation operators in $T$ and in $D$, respectively.
Analogously, $A_T,A_D$ are the clusters of their respective annihilation operators. 
Normal-ordering the product $TD$ can cancel local operators only on the intersection
\begin{align}
 S_{TD}=A_T \cap C_D.
\end{align}
Due to the locality of the triplon algebra, the commutator vanishes if none of the clusters overlap
\begin{align}
 S_{TD}=\emptyset \wedge S_{DT}=\emptyset.
\end{align}
Thus the normal-ordered product $TD$ definitely has local creation operators on the union cluster
\begin{subequations}
 \begin{align}
 	\label{eq:C_min_set}
  C_{TD}\supseteq C_T\cup\left(C_D\setminus S_{TD}\right)
 \end{align}
and local annihilation operators on the union cluster
 \begin{align}
 	\label{eq:A_min_set}
  A_{TD}\supseteq A_D\cup\left(A_T\setminus S_{TD}\right) .
 \end{align}
\end{subequations}
There may be additional creation or annihilation operators, but no general statements
can be made on their existence. In this sense, the right-hand sides of
Eqs.\ \eqref{eq:C_min_set} and \eqref{eq:A_min_set} are minimum clusters for the normal-ordered product
$TD$. They can be used in Eq.\ \eqref{eq:omaxbound_general} to obtain an upper bound for the maximum order ${{\widetilde{O}}_{\max}}{}_{,TD}$ and the corresponding reasoning is used to obtain ${{\widetilde{O}}_{\max}}{}_{,DT}$.
This makes it possible to avoid the computation of the commutator $[ T,D ] $.

Moreover, one can use the intersections $S_{TD}$ and $S_{DT}$ to exploit the hard-core property of the triplons: The normal-ordered product $TD$ will vanish if $C_T$ and $C_D\setminus S_{TD}$ are not disjoint or likewise if $A_D$ and $\left(A_T\setminus S_{TD}\right)$ are not disjoint because the creation or
annihilation of two triplons is attempted on the same site.

Although it is less strict, the basic \emph{a priori} rule has the advantage to be much more lightweight in comparison to the extended \emph{a priori} rule because
it requires mere counting of operators. Furthermore, it can be used very efficiently in the context of translation symmetry. Because it does not rely on the real-space structure of a term, it can be applied to all terms in the translation group in contrast to the extended rule. Therefore, for best performance, it turns out to be most efficient to combine both rules in practice.

\begin{table*}
\section{Series expansion of ground-state energy}
 \begin{tabular}{|d||d|d|d|d|d|d|}
\hline
\multicolumn{1}{|c||}{order} & \multicolumn{1}{c|}{$y=1$} & \multicolumn{1}{c|}{$y=1.1$} & \multicolumn{1}{c|}{$y=1.2$} & \multicolumn{1}{c|}{$y=1.3$} & \multicolumn{1}{c|}{$y=1.4$} & \multicolumn{1}{c|}{$y=1.5$} \\
\hline
0 & -0.75 & -0.7875 & -0.825 & -0.8625 & -0.9 & -0.9375 \\
1 & 0 & 0 & 0 & 0 & 0 & 0 \\
2 & -0.37500000 & -0.35714286 & -0.34090909 & -0.32608696 & -0.31250000 & -0.30000000 \\
3 & -0.18750000 & -0.17006803 & -0.15495868 & -0.14177694 & -0.13020833 & -0.12000000 \\
4 & 0.02343750 & 0.02006213 & 0.01702197 & 0.01434365 & 0.01201326 & 0.01000000 \\
5 & 0.17578125 & 0.14444038 & 0.11952736 & 0.09957578 & 0.08347930 & 0.07040000 \\
6 & 0.15527344 & 0.12203681 & 0.09750810 & 0.07902742 & 0.06485370 & 0.05381334 \\
7 & -0.05364990 & -0.03921398 & -0.02800412 & -0.01955753 & -0.01330019 & -0.00871467 \\
8 & -0.27630616 & -0.19638549 & -0.14184565 & -0.10397006 & -0.07724516 & -0.05811099 \\
9 & -0.23688412 & -0.16217018 & -0.11516891 & -0.08425661 & -0.06315254 & -0.04828763 \\
10 & 0.16046858 & 0.10149005 & 0.06334061 & 0.03903035 & 0.02365644 & 0.01398225 \\
11 & 0.58532060 & 0.36106918 & 0.22967669 & 0.15006081 & 0.10038628 & 0.06858346 \\
12 & 0.43494050 & 0.26028493 & 0.16564720 & 0.11039861 & 0.07614142 & 0.05388099 \\
13 & -0.50265913 & -0.27674749 & -0.15351333 & -0.08578336 & -0.04820240 & -0.02714962 \\
14 & -1.41593560 & -0.75955267 & -0.42752222 & -0.25046274 & -0.15174318 & -0.09458886 \\
15 & -0.84414414 & -0.44261065 & -0.25354303 & -0.15442048 & -0.09802426 & -0.06402580 \\
16 & 1.60970122 & 0.77306865 & 0.38294840 & 0.19499667 & 0.10171476 & 0.05416803 \\
17 & 3.67381373 & - & - & - & - & -\\
\hline
\end{tabular}
\caption{Coefficients of the perturbative evaluation for the ground-state energy per rung for various $y=\nicefrac{J_{\bot}^{\text{o}}}{J_{\bot}^{\text{e}}}$. 
\label{tab:coefficients_gse}}
\end{table*}

\begin{table*}
\section{Series expansion of hopping terms}
\scalebox{0.8}{
 \begin{tabular}{|d||d|d|d||d|d|}
\hline
\multicolumn{1}{|c||}{order}  & \multicolumn{1}{c|}{$y=1$} &  \multicolumn{1}{c|}{$y=1.2\,(M_{\text{ee}})$} &  \multicolumn{1}{c||}{$y=1.2\,(M_{\text{oo}})$} &  \multicolumn{1}{c|}{$y=1$} &  \multicolumn{1}{c|}{$y=1.2\,(M_{\text{eo}})$} \\
\hline
& \multicolumn{3}{c||}{$t_0$} & \multicolumn{2}{c|}{$t_1$} \\
\hline
0  & 1           & 1           &  1.2        & -           & - \\
1  & 0           & 0           &  0          & 0.5         & 0.5 \\
2  & 0.75000000  &  0.68181818 &  0.68181818 & 0           & 0 \\
3  & 0.37500000  &  0.30991736 &  0.30991736 & -0.12500000 & -0.10351967 \\
4  & -0.20312499 & -0.14261905 & -0.15986342 & -0.15625000 & -0.11787954 \\
5  & -0.62500000 & -0.41557085 & -0.43722511 & -0.10156250 & -0.07018606 \\
6  & -0.50000000 & -0.30923490 & -0.32459462 & 0.04687500  &  0.02853343 \\
7  & 0.29663086  &  0.15536931 &  0.15988661 & 0.16467285  &  0.09328777 \\
8  & 1.12030030  &  0.56447272 &  0.59149035 & 0.12779236  &  0.06680455 \\
9  & 0.90001680  &  0.42714378 &  0.46150160 & -0.08070850 & -0.03759727 \\
10 & -0.75448108 & -0.30571050 & -0.29283265 & -0.24961996 & -0.10854657 \\
11 & -2.44631335 & -0.95489311 & -0.98116281 & -0.08650172 & -0.03556142 \\
12 & -1.60154019 & -0.60166396 & -0.65043831 & 0.41219152  &  0.15043006 \\
13 & 2.59697176  &  0.82161561 &  0.79729874 & 0.69643046  &  0.23728904 \\
14 & 6.30730682  & - & - & 0.03314883 & - \\
15 & 2.83300346  & - & - & -1.4674947 & - \\
\hline
& \multicolumn{3}{c||}{$t_2$} & \multicolumn{2}{c|}{$t_3$} \\
\hline
2  & -0.12500000 & -0.11363636 & -0.11363636 & - & - \\
3  & -0.12500000 & -0.10330579 & -0.10330579 & 0.06250000  &  0.05175983 \\
4  & -0.01562500 & -0.01227400 & -0.01120458 & 0.06250000  &  0.04715182 \\
5  & 0.10156250  &  0.06924276 &  0.06922755 & -0.04687500 & -0.03205290 \\
6  & 0.08593750  &  0.05785942 &  0.04855859 & -0.15820313 & -0.09861260 \\
7  & -0.08642578 & -0.04090435 & -0.05704574 & -0.11114502 & -0.06343140 \\
8  & -0.25237274 & -0.12952772 & -0.13188402 &  0.13763428 &  0.06993628 \\
9  & -0.13074875 & -0.07813024 & -0.04833601 &  0.37034416 &  0.17332712 \\
10 & 0.34451961  &  0.12460577 &  0.16808751 &  0.21864462 &  0.09568796 \\
11 & 0.73779087  &  0.28987133 &  0.29425793 & -0.40222562 & -0.15201914 \\
12 & 0.33147282  &  0.15944178 &  0.09654811 & -0.90268454 & -0.31908032 \\
13 & -0.99486640 & -0.27885442 & -0.35491362 & -0.33963371 &  -0.11778655 \\
14 & -1.98783536 & - & - & 1.36473269 & - \\
15 & -0.70791841 & - & - & 2.41431850 & - \\
\hline
& \multicolumn{3}{c||}{$t_4$} & \multicolumn{2}{c|}{$t_5$} \\
\hline
4  & -0.03906250 & -0.03116138 & -0.02792639 & - & - \\
5  & -0.04687500 & -0.03317747 & -0.03121078 &  0.02734375 &  0.01880795 \\
6  &  0.03564453 &  0.02548558 &  0.01949955 &  0.03906250 &  0.02443938 \\
7  &  0.13452148 &  0.08115556 &  0.07210783 & -0.03021240 & -0.01723325 \\
8  &  0.08452606 &  0.04273061 &  0.04437608 & -0.13285828 & -0.06882735 \\
9  & -0.16891479 & -0.08803878 & -0.07128815 & -0.09355831 & -0.04418694 \\
10 & -0.37874413 & -0.17065003 & -0.15393594 &  0.18216133 &  0.07772144 \\
11 & -0.12521664 & -0.04650305 & -0.05271784 &  0.44106736 &  0.17211308 \\
12 &  0.63520241 &  0.24143238 &  0.20728113 &  0.16840086 &  0.06173547 \\
13 &  1.09758909 &  0.37362081 &  0.34061615 & -0.74875764 & -0.23909940 \\
14 &  0.10661141 & - & - & -1.34708600 & - \\
15 & -2.19284584 & - & - & -0.12406771 & - \\
\hline
& \multicolumn{3}{c||}{$t_6$} & \multicolumn{2}{c|}{$t_7$} \\
\hline
6  & -0.02050781 & -0.01398812 & -0.01184453 & - & - \\
7  & -0.03417969 & -0.02091311 & -0.01819031 &  0.01611328 &  0.00921603 \\
8  &  0.02478790 &  0.01521149 &  0.01095480 &  0.03076172 &  0.01599404 \\
9  &  0.13085937 &  0.06817047 &  0.05621627 & -0.02009487 & -0.00957393 \\
10 &  0.10456268 &  0.04648122 &  0.04328735 & -0.12947965 & -0.05581397 \\
11 & -0.19036049 & -0.08638434 & -0.06396735 & -0.11682585 & -0.04574993 \\
12 & -0.50204569 & -0.19572116 & -0.16328447 &  0.19740278 &  0.07034140 \\
13 & -0.20806881 & -0.06412749 & -0.07130687 &  0.57210467 &  0.18566781 \\
14 &  0.90379496 & - & - &  0.26874640 & - \\
15 &  1.68652741 & - & - & -1.06182222 & - \\
\hline
& \multicolumn{3}{c||}{$t_8$} & \multicolumn{2}{c|}{$t_9$} \\
\hline
8  & -0.01309204 & -0.00755457 & -0.00618286 & - & - \\
9  & -0.02819824 & -0.01468727 & -0.01219007 &  0.01091003 &  0.00519218  \\
10 &  0.01595318 &  0.00838267 &  0.00574076 &  0.02618408 &  0.01132456 \\
11 &  0.12834901 &  0.05682685 &  0.04493297 & -0.01227187 & -0.00490073 \\
12 &  0.13062643 &  0.05022786 &  0.04326984 & -0.12730413 & -0.04569697 \\
13 & -0.20082120 & -0.07779488 & -0.05499657 & -0.14531211 & -0.04729919 \\
14 & -0.64685442 & - & - &  0.20129516 & - \\
15 & -0.34698491 & - & - &  0.72601478 & - \\
\hline
& \multicolumn{3}{c||}{$t_{10}$} & \multicolumn{2}{c|}{$t_{11}$} \\
\hline
10 & -0.00927353 & -0.00450284 & -0.00360404 & - & - \\
11 & -0.02454758 & -0.01079578 & -0.00869977 &  0.00800896 &  0.00317279 \\
12 &  0.00896406 &  0.00409773 &  0.00260778 &  0.02318382 &  0.00834536 \\
13 &  0.12625622 &  0.04714879 &  0.03633844 & -0.00596249 & -0.00201914 \\
14 &  0.16070462 & - & - & -0.12514982 & - \\
15 & -0.19870083 & - & - & -0.17663111 & - \\
\hline
& \multicolumn{3}{c||}{$t_{12}$} & \multicolumn{2}{c|}{$t_{13}$} \\
\hline
12 & -0.00700784 & -0.00285462 & -0.00224969 & - & - \\
13 & -0.02202463 & -0.00814201 & -0.00643333 & 0.00619924 &  0.00204486 \\
14 &  0.00321460 & - & - &  0.02102351 & - \\
15 &  0.12395082 & - & - & -0.00067934 & - \\
\hline
& \multicolumn{3}{c||}{$t_{14}$} & \multicolumn{2}{c|}{$t_{15}$} \\
\hline
14 & -0.00553504 & - & - & - & - \\
15 & -0.02014753 & - & - &  0.00498153 & - \\
\hline
\end{tabular}
}
\caption{Coefficients of the perturbative evaluation for the dispersion for $y=1$ and $y=1.2$. \label{tab:coefficients_dispersion}}
\end{table*}

\end{document}